\documentclass[10pt,conference]{IEEEtran}
\usepackage{cite}
\usepackage{amsmath,amssymb,amsfonts}
\usepackage{algorithmic}
\usepackage{graphicx}
\usepackage{textcomp}
\usepackage{xcolor}
\usepackage[hyphens]{url}
\usepackage{fancyhdr}
\usepackage{hyperref}
\usepackage[normalem]{ulem}

\newcommand{\hpcayear}{2025}

\newcommand{\hpcasubmissionnumber}{NaN}

\newcommand{\Arch}{EXION}
\newcommand{\agthmsFFN}{FFN-Reuse}
\newcommand{\agthmsCD}{ConMerge}

\title{\Arch: Exploiting Inter- and Intra-Iteration Output Sparsity for Diffusion Models}

\def\hpcacameraready{} 

\newcommand\hpcaauthors{Jaehoon Heo, Adiwena Putra, Jieon Yoon, Sungwoong Yune, Hangyeol Lee, Ji-Hoon Kim, and Joo-Young Kim}
\newcommand\hpcaaffiliation{\textit{KAIST}\\\textit{Daejeon, Republic of Korea}}
\newcommand\hpcaemail{\{kd01050, adiwena.research, j9e8y, imwooong, lhg4294, jihoon0708, jooyoung1203\}@kaist.ac.kr}

\author{
  \ifdefined\hpcacameraready
    \IEEEauthorblockN{\hpcaauthors{}}
      \IEEEauthorblockA{
        \hpcaaffiliation{} \\
        \hpcaemail{}
      }
  \else
    \IEEEauthorblockN{\normalsize{HPCA \hpcayear{} Submission
      \textbf{\#\hpcasubmissionnumber{}}} \\
      \IEEEauthorblockA{
        Confidential Draft \\
        Do NOT Distribute!!
      }
    }
  \fi 
}

\fancypagestyle{camerareadyfirstpage}{%
  \fancyhead{}
  
  \fancyhead[C]{
    \ifdefined\aeopen
    \parbox[][12mm][t]{13.5cm}{\hpcayear{} IEEE International Symposium on High-Performance Computer Architecture (HPCA)}    
    \else
      \ifdefined\aereviewed
      \parbox[][12mm][t]{13.5cm}{\hpcayear{} IEEE International Symposium on High-Performance Computer Architecture (HPCA)}
      \else
      \ifdefined\aereproduced
      \parbox[][12mm][t]{13.5cm}{\hpcayear{} IEEE International Symposium on High-Performance Computer Architecture (HPCA)}
      \else
      \parbox[][0mm][t]{13.5cm}{\hpcayear{} IEEE International Symposium on High-Performance Computer Architecture (HPCA)}
    \fi 
    \fi 
    \fi 
    \ifdefined\aeopen 
      \includegraphics[width=12mm,height=12mm]{ae-badges/open-research-objects.pdf}
    \fi 
    \ifdefined\aereviewed
      \includegraphics[width=12mm,height=12mm]{ae-badges/research-objects-reviewed.pdf}
    \fi 
    \ifdefined\aereproduced
      \includegraphics[width=12mm,height=12mm]{ae-badges/results-reproduced.pdf}
    \fi
  }
  \fancyfoot[C]{}
}
\begin{document}
\maketitle

\ifdefined\hpcacameraready 
  \thispagestyle{camerareadyfirstpage}
  \pagestyle{empty}
\else
  \thispagestyle{plain}
  \pagestyle{plain}
\fi

\newcommand{\hpcaheight}{0mm}
\ifdefined\eaopen
\renewcommand{\hpcaheight}{12mm}
\fi

\begin{abstract}

Over the past few years, diffusion models have emerged as novel solutions in AI industries, offering the capability to generate diverse, multi-modal outputs such as images, videos, and motions from input prompts (i.e., text). Despite their impressive capabilities, diffusion models face significant challenges in computing, including excessive latency and energy consumption due to the numerous iterations in model architecture. Although prior works specialized in transformer acceleration can be applied to diffusion models, given that transformers are key components, the problem of the iterative nature of diffusion models still needs to be addressed. 

In this paper, we present EXION, the first software-hardware co-designed diffusion accelerator that solves the computation challenges of excessive iterations by exploiting the unique inter- and intra-iteration output sparsity in diffusion models. To this end, we propose two software-level optimizations in EXION. First, we introduce the FFN-Reuse algorithm that identifies and skips redundant computations in FFN layers across different iterations (i.e., inter-iteration sparsity). Second, we use a modified eager prediction method that employs two-step leading-one detection to accurately predict the attention score in diffusion models, skipping unnecessary computations within an iteration (i.e., intra-iteration sparsity). We also introduce a novel data compaction mechanism named ConMerge, which can enhance hardware utilization by condensing and merging large and sparse matrices into small and compact forms. Finally, EXION has a dedicated hardware architecture that supports the above sparsity-inducing algorithms, translating high output sparsity into improved energy efficiency and performance.
To verify the feasibility of the EXION accelerator, we first demonstrate that it has no impact on accuracy in various types of multi-modal diffusion models, including text-to-motion, -audio, -image, and -video. We then instantiate EXION in both server- and edge-level settings and compare its performance against GPUs with similar specifications. Our evaluation shows that EXION achieves dramatic improvements in performance and energy efficiency by 3.2-379.3$\times$ and 45.1-3067.6$\times$ compared to a server GPU and by 42.6-1090.9$\times$ and 196.9-4668.2$\times$ compared to an edge GPU.

\end{abstract}
\section{Introduction}
\label{section1}

Recently, the demand for artificial intelligence (AI) that can imagine and create innovative information like humans has been a driving force in machine learning (ML) research. As a result, diffusion models~\cite{ho2020denoising} have emerged as promising solutions in various generative AI industries. For instance, with a simple text prompt as input, Stable Diffusion~\cite{rombach2022high} can generate high-quality images, and Sora~\cite{videoworldsimulators2024} can generate vivid and complex videos. Some models can easily generate human motions from text~\cite{tevet2023human} or audio~\cite{tseng2023edge} inputs. This versatility highlights their applicability across multiple applications, making diffusion models a crucial aspect of generative AI technology.

Although diffusion models can create remarkably accurate and detailed outputs, they come with high energy consumption and long latency. This is due to their fundamental characteristic of starting with random noise and progressively removing it to obtain the desired outputs over numerous iterations, which can reach up to 1000~\cite{ho2020denoising}. Furthermore, as diffusion models are primarily based on transformer blocks~\cite{vaswani2017attention}, each iteration involves a significant number of operations. Consequently, their energy consumption and latency are significantly higher than those of other generative models. For example, Stable Diffusion generates images from text prompts with better quality and diversity compared to StyleGAN-XL~\cite{sauer2022stylegan}, a popular generative adversarial network. However, our experimental results show that on an identical GPU device, NVIDIA's RTX 6000 Ada~\cite{rtx6000}, the energy consumption and latency of Stable Diffusion are 1546.7J and 11.8s, respectively, which are 23.6$\times$ and 9.8$\times$ higher than those of StyleGAN-XL.

To solve these problems derived from repeated iterations, a few software (SW)-based approaches~\cite{salimans2022progressive, chung2022come, lu2022dpm, song2023consistency} that optimize the scheduler of the diffusion model have been proposed to reduce the large number of iterations, i.e., inference steps. However, these fast sampling methods come at the cost of accuracy~\cite{zheng2023fast}. Additionally, some of them even require retraining of the pre-trained diffusion model~\cite{luo2023lcm}.

Another approach to accelerating the diffusion model is using specialized hardware (HW) accelerators to speed up a large number of operations in each iteration without modifying the scheduler aggressively. Existing transformer accelerators can be utilized~\cite{qin2023fact, lu2021sanger, ham2021elsa} to speed up computations in each iteration. Most previous transformer accelerators focus on optimizing the query, key, and value (QKV) projection and attention computation using methods such as eager prediction~\cite{qin2023fact} and attention pruning~\cite{lu2021sanger}. However, these optimizations do not significantly reduce the energy consumption and latency of the overall diffusion process since they have not exploited the iterative nature of it. To achieve optimal performance, it is necessary to exploit the unique sparsity that exists only in diffusion models.

\begin{figure}[t]
\centering
\includegraphics[width=3.4in]{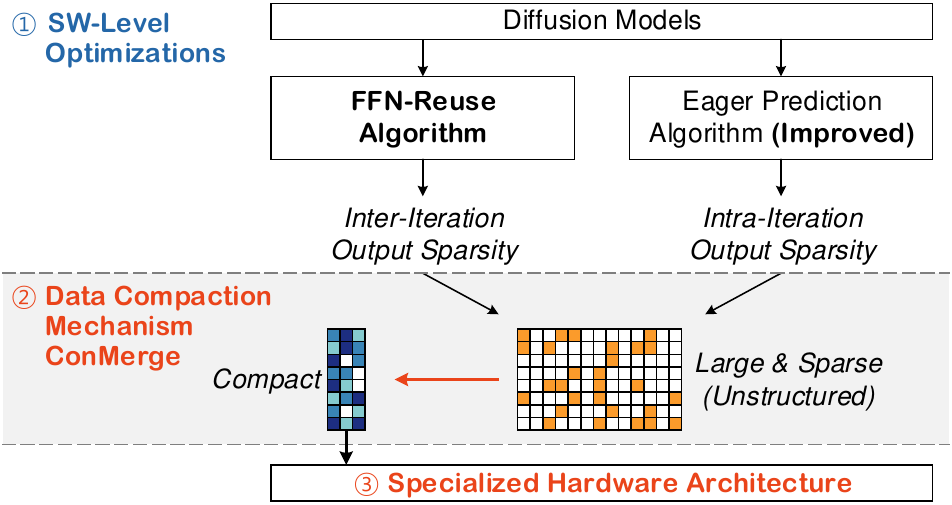}
\caption{Overview of {\Arch} Accelerator}
\label{fig1}
\vspace{-0.1in}
\end{figure}

In this work, we present {\Arch}, the first diffusion accelerator developed to solve the energy consumption and latency problems in diffusion models by exploiting inter- and intra-iteration output sparsity. To this end, we adopt a SW-HW co-design strategy, achieving both output sparsities through SW-level optimizations and fully utilizing them through a data compaction mechanism and dedicated architecture.

Figure~\ref{fig1} shows the overview of {\Arch}. At the SW-level optimization, {\Arch} introduces the {\agthmsFFN} algorithm, which identifies data redundancy in the matrix multiplication (MMUL) outputs of feedforward neural network (FFN) layers in transformer blocks across different iterations. By utilizing this unique property, the {\agthmsFFN} algorithm reuses the previous iteration's output data for several subsequent iterations, eliminating the need to compute the inner product for those particular elements and making some output data to be sparse. We refer to this phenomenon as inter-iteration output sparsity. Specifically, it differs from the well-known weight or input sparsities resulting from methods like structured pruning or activation functions, as it arises from data reuse in the FFN's output and can occur independently of them.
Moreover, within the execution of a single iteration, {\Arch} has improved the eager prediction algorithm and made it applicable to diffusion models. Similar to inter-iteration output sparsity, the improved method leads to another type of output sparsity during MMUL within the same iteration and the same transformer block, which we term intra-iteration output sparsity.

Although the percentage of resulted output sparsity can dramatically reach up to 97\% in our benchmarks, conventional HW, such as GPUs, cannot reduce energy consumption and latency by utilizing it due to the unstructured sparsity in the output matrix. Since GPUs can only manage sparsity in a coarse-grained manner, they are not effective at handling fine-grained sparsity. To address this issue, we propose a data compaction mechanism named {\agthmsCD} that enhances HW utilization by condensing and merging large and sparse matrices into small and compact forms. Finally, we developed a tailored architecture for this mechanism, executing operations on the compact matrix forms. Notably, this architecture can handle output sparsity with a flexible yet simple datapath, unlike other accelerators~\cite{qin2020sigma,yangtrapezoid}, which suffer additional overheads for data feeding to handle sparsity in input or weight data. Additionally, {\Arch} includes an assistance unit to generate control signals for the {\agthmsCD} mechanism. Consequently, it successfully translates high output sparsity into improvements in energy efficiency and performance.

To verify the efficiency and versatility of {\Arch}, we first demonstrate that its accuracy loss is trivial across various types of diffusion models. Finally, we implemented a cycle-level simulator and compared its performance against the NVIDIA RTX 6000 Ada GPU, verifying that {\Arch}'s performance and energy efficiency increased by 3.2-379.3$\times$ and 45.1-3067.6$\times$, respectively. We summarize the contributions of this work as follows:

\begin{itemize}
    \item We propose {\Arch}, the first software-hardware co-designed diffusion accelerator that solves the computation challenges of excessive iterations by exploiting the unique inter- and intra-iteration output sparsity in diffusion models.
    \item At the SW-level optimizations, {\Arch} focuses on achieving output sparsity. It introduces the {\agthmsFFN} algorithm for unique inter-iteration sparsity in FFN layers and proposes an accurate eager prediction method to achieve intra-iteration sparsity in attention computations.
    \item To efficiently manage sparsity, {\Arch} incorporates the data compaction mechanism {\agthmsCD}, which condenses and merges large, sparse matrices into smaller, more compact forms.
    \item Finally, {\Arch} is equipped with a specialized HW architecture that effectively handles the {\agthmsCD} mechanism, resulting in significant improvements in performance and energy efficiency compared to both server- and edge-level GPUs.
\end{itemize}
\section{Background and Motivation}
\label{section2}

\begin{figure}[t]
\centering
\includegraphics[width=3.4in]{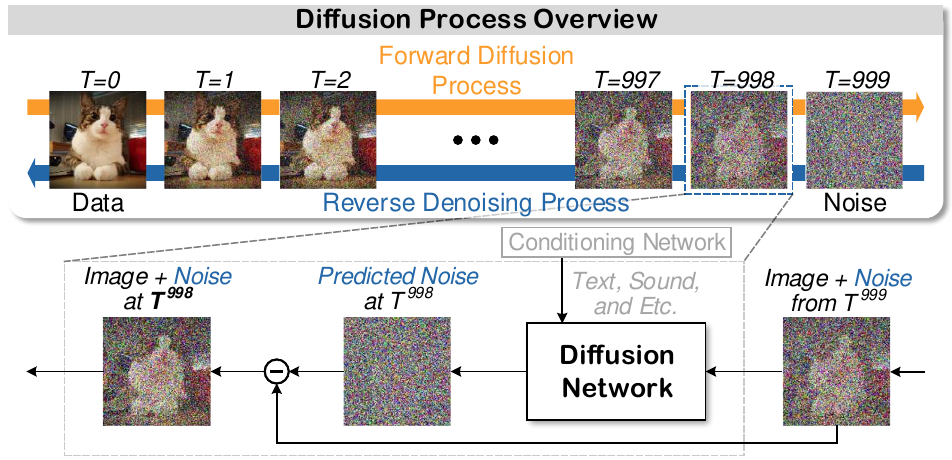}
\caption{Overview of Diffusion Models}
\label{fig2}
\vspace{-0.1in}
\end{figure}

\subsection{Overview of Diffusion Models}
\label{section2_1}

Figure~\ref{fig2} shows an overview of the diffusion process, which comprises forward diffusion and reverse denoising processes, using a text-to-image generation example~\cite{nvidia_diffusion_presentation_cvpr}. During the forward diffusion process, the model progressively adds noise to the original input data over numerous iterations, converting the given input to noise. In contrast, the reverse denoising process starts from noise and revives the original input data.

During the training phase, the diffusion network learns to predict noise from input data at the corresponding iteration while repeating the forward diffusion and reverse denoising processes. By subtracting the predicted noise from the input data, the network produces output data with a reduced noise level, which becomes the input for the next iteration. For instance, as shown in the figure, at iteration 998 (i.e., when T is 998), the diffusion network predicts the noise at T$^{998}$ using the input received from the previous iteration. Then, it subtracts the predicted noise from that input, resulting in a cleaner image, which will be the input for the next iteration, T$^{997}$. During the inference phase, only the reverse denoising process is conducted to generate output data from noise. In this paper, we primarily focus on reducing energy consumption and execution time during the inference phase.

When there are conditional inputs, such as text or sound, transformer-based conditioning networks like CLIP~\cite{radford2021learning} and CLAP~\cite{elizalde2023clap} are executed once to convert these inputs into embeddings. These embeddings are then integrated into the diffusion network during its denoising steps, allowing the generated output to reflect the provided inputs accurately.

\begin{figure}[t]
\centering
\includegraphics[width=3.4in]{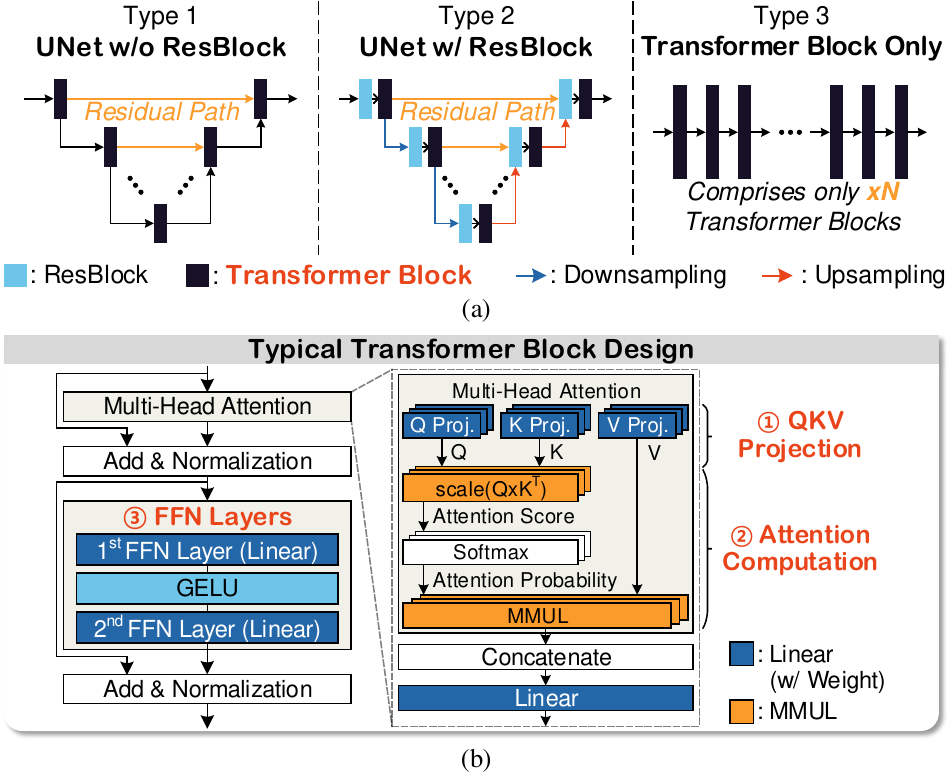}
\caption{(a) Diffusion Network and (b) Transformer Block Design}
\label{fig3}
\vspace{-0.1in}
\end{figure}

As shown in Figure~\ref{fig3} (a), the most important block in the diffusion network is the transformer block, which is included in all types of diffusion models and typically has a large number of operations. Figure~\ref{fig3} (b) illustrates a typical design of the transformer block, where the projection of the query (Q), key (K), and value (V) is first conducted by performing MMUL between the given input and the respective weights. Second, the attention computation is computed by executing MMUL between Q and K$^T$, applying the Softmax function, and then performing MMUL with V. Finally, after concatenating all the multi-head attention outputs, the FFN layers are executed.

\begin{figure}[t]
\centering
\includegraphics[width=3.4in]{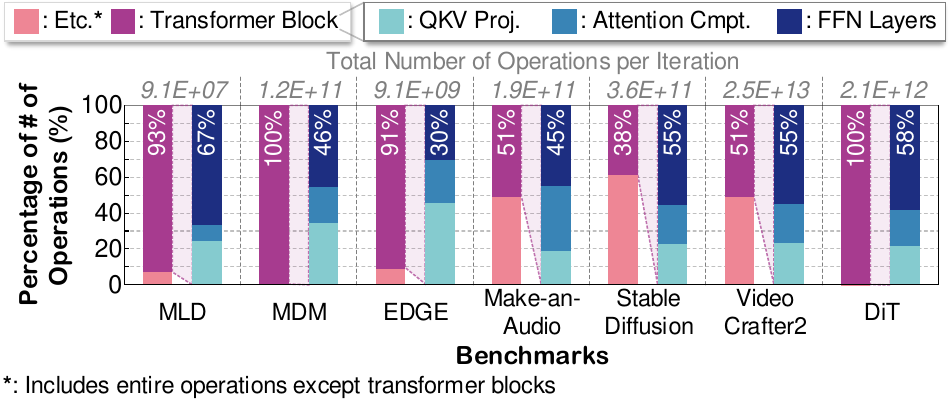}
\caption{Number of Operations Breakdown}
\label{fig4}
\end{figure}

Figure~\ref{fig4} depicts the breakdown of the number of operations in our benchmark models. It demonstrates that the transformer block, including QKV projection, attention computation, and FFN layers, accounts for the highest ratio of operations, ranging from 38\% to 100\%, leading to the highest energy consumption. Notably, within the transformer block, the FFN layers are generally the most compute-intensive. This is because diffusion models have relatively short token lengths compared to recent large language models, which have significantly longer token lengths and thus require much more computation for QKV projection and attention computation. As a result, as shown in the figure, FFN layers become the main bottleneck with the highest computational overhead in the transformer block of diffusion models, reaching up to 67\%. Although prior transformer accelerators have suggested optimization methods, they mostly target the QKV projection and attention computation and are not ideal for the FFN layers in diffusion models. Consequently, we propose a novel method to address the significant overhead of FFN layers in diffusion models, which is described in Section~\ref{section3}.

\subsection{Eager Prediction Algorithm}
\label{section2_2}

\begin{figure}[t]
\centering
\includegraphics[width=3.4in]{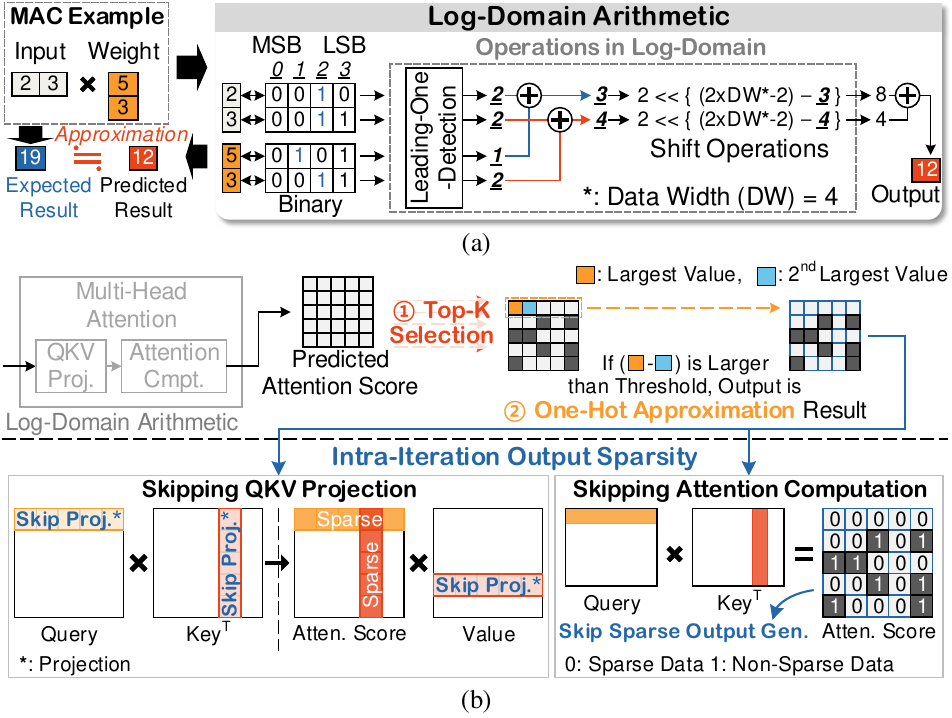}
\caption{(a) Log-Domain Arithmetic in EP (b) Intra-Iteration Output Sparsity}
\label{fig5}
\vspace{-0.1in}
\end{figure}

While the number of operations in FFN layers is dominant, this does not mean that QKV projection and attention computation are trivial in diffusion models. To address these efficiently, we adopt an eager prediction (EP) algorithm from the state-of-the-art (SOTA) transformer accelerator~\cite{qin2023fact}. This algorithm first predicts the attention score using simple log-domain arithmetic. Then, using the predicted attention score, it identifies unnecessary output results that can be skipped in more complex real-domain computations.

Figure~\ref{fig5} (a) explains the log-domain arithmetic in detail. It begins with preprocessing by separating the input and weight data of QKV projection into sign bits and absolute values. Then, it identifies the position of the leading-one bit, i.e., the first bit that is set to one from the most significant bit (MSB), effectively approximating an integer number to a log-domain number. In multiply-and-accumulate (MAC) operations, the multiplication is simplified to an addition followed by a shift operation that converts the log-domain number back to a real-domain number. Finally, it conducts the remaining additions for accumulation, considering the sign bit information.

This straightforward log-domain arithmetic also applies to attention computation to generate the predicted attention score. Since the predicted score serves as the input to the Softmax function, the algorithm leverages the characteristic of the function, where non-important elements are always present in a row of Softmax outputs. As illustrated in Figure~\ref{fig5} (b), we initially perform a top-k selection on each row of the predicted attention score. Values that do not rank within the "top k" are directly assigned to zero, as their influence will be negligible in the subsequent layer after the Softmax function. Moreover, when the difference between the most dominant value and the second most dominant value surpasses a predetermined threshold, the computation for that row can be entirely skipped. This is because the dominant element is already determined, rendering the remaining elements in that row effectively zero. Finally, skipping computation on the negligible elements in the predicted attention score will lead to output sparsity within each iteration during the attention computation, which we term as intra-iteration output sparsity.

As a result, the EP technique achieves intra-iteration output sparsity in attention computation ranging from 20\% to 95\% in our benchmarks. Additionally, another merit of EP is that when all the elements in a column of the attention score are sparse, we can even skip K and V projection for those columns. Furthermore, when the output is generated directly through one-hot approximation due to one dominant value in a row, the Q projection for that row can be skipped. Finally, in our benchmarks, the EP algorithm skips, on average, 26\% of Q projection and 22\% of KV projection.

However, the EP method cannot be directly applied to diffusion models, where various transformer blocks exist within a single iteration, and furthermore, those iterations are continuously repeated up to 1000 times. The error due to data approximation accumulates, leading to decreased accuracy. Therefore, we have refined this method by increasing its prediction accuracy to make it applicable to diffusion models. The detailed improvements are delineated in Section~\ref{section4_4}.
\section{{\Arch's} Software-Hardware Co-Design Strategy}
\label{section3}

To reduce the overhead of numerous operations resulting from excessive iterations, {\Arch} first provides SW-level optimizations consisting of the {\agthmsFFN} and EP algorithms, which lead to inter- and intra-iteration output sparsity patterns, respectively. Notably, the main novelty of this work and the resulting efficiency gains come from the {\agthmsFFN} algorithm. This algorithm exploits the iterative nature of the diffusion model, achieving unique inter-iteration output sparsity in FFN layers, which are the main overhead of the transformer block. In this section, we will focus on the major aspects that {\Arch} brings, particularly the {\agthmsFFN} algorithm, followed by the data compaction mechanism, {\agthmsCD}.

\subsection{{\agthmsFFN} Algorithm for Inter-Iteration Output Sparsity}
\label{section3_1}

\begin{figure}[t]
\centering
\includegraphics[width=3.4in]{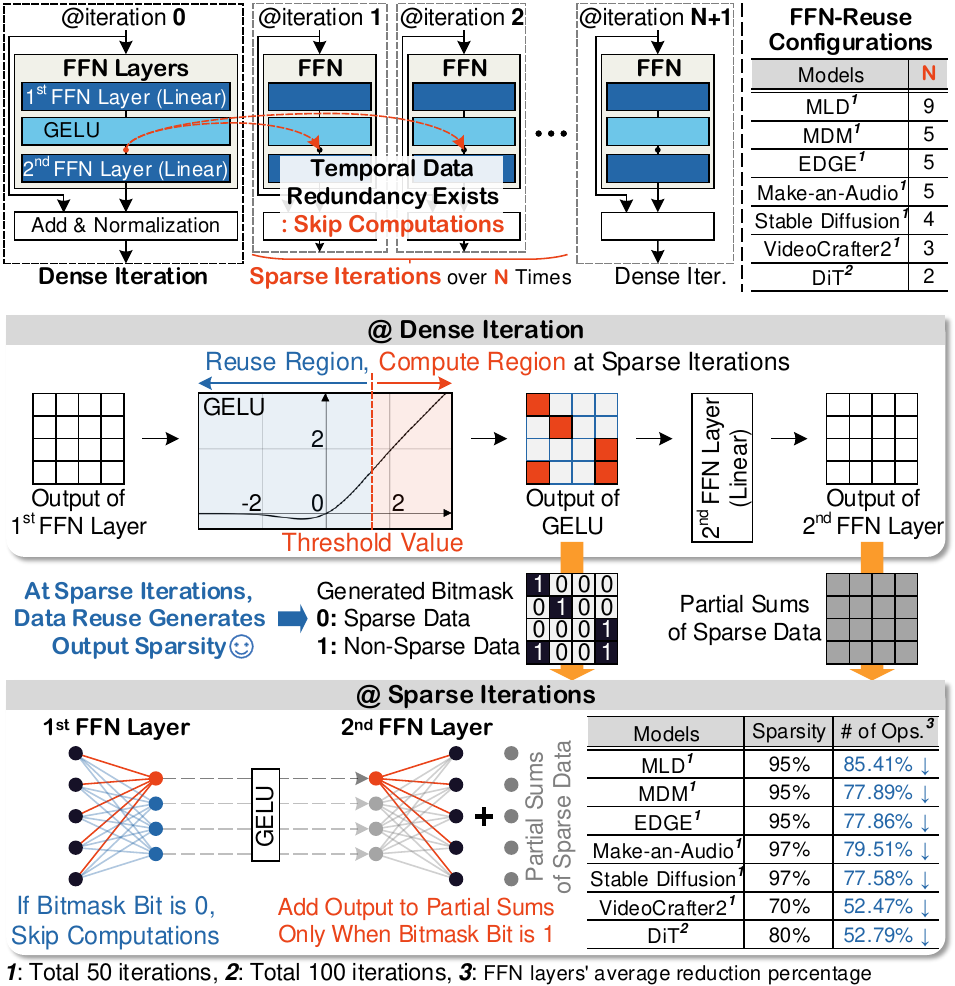}
\caption{{\agthmsFFN} Algorithm for Inter-iteration Output Sparsity}
\label{fig6}
\vspace{-0.1in}
\end{figure}

Focusing on the fundamental characteristic of diffusion models that progressively removes noise over numerous iterations, we examined data patterns across these iterations. As a result, we discovered that temporal data redundancy exists across different iterations in FFN layers. Consequently, we propose a {\agthmsFFN} algorithm that can significantly reduce the number of operations in FFN layers by skipping redundant computations.

Figure~\ref{fig6} describes the algorithm in detail. We first conduct the normal diffusion, i.e., computing all operations required in the FFN layers for a single iteration, which we call a dense iteration. In dense iteration, we also generate the bitmask according to the output of the non-linear layer (e.g., GELU~\cite{hendrycks2016gaussian} or GEGLU~\cite{shazeer2020gluvariantsimprovetransformer}) by comparing the output values with a predetermined threshold. Values larger than this threshold are considered important and need to be recomputed at every iteration, while values smaller than the threshold will be reused for the next \textit{N} iterations, which we refer to as sparse iterations.

The table in Figure~\ref{fig6} shows that the output sparsity of the 1st FFN layer achieves 70\% to 97\% on average. The proposed {\agthmsFFN} leverages this high output sparsity, directly skipping the 1st FFN layer's unnecessary computations. The 2nd FFN layer also skips unnecessary computations by only accumulating the updated values to a partial sum derived from dense iteration. As a result, by continuously executing one dense iteration and \textit{N} sparse iterations during the entire diffusion process, we can skip 52.47\% to 85.41\% of operations in FFN layers within our benchmarks.

\begin{figure}[t]
\centering
\includegraphics[width=3.4in]{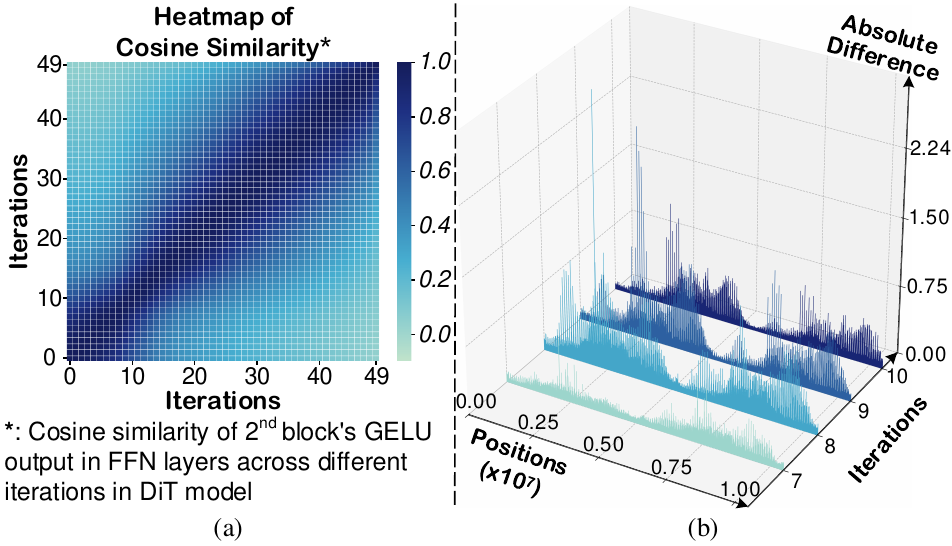}
\caption{(a) Heatmap of Cosine Similarity (b) Difference Between Adjacent Iterations}
\label{fig7}
\vspace{-0.1in}
\end{figure}

The rationale behind why {\agthmsFFN} is a feasible solution can be found by analyzing the characteristics of diffusion models. Figure~\ref{fig7} (a) shows the cosine similarity across different iterations in the DiT model~\cite{peebles2023scalable}. Specifically, it compares the output of the GELU function that exists between two FFN layers. The figure indicates that the cosine similarity between adjacent iterations is high, suggesting the potential for temporal data reuse.

Although most of the outputs can be reused across different iterations, some of the output data cannot be reused. Figure~\ref{fig7} (b) shows the difference in GELU outputs between adjacent iterations. While most of the data shows trivial differences, some points have large differences, indicating computation is required at these points. Notably, the positions where large differences occur are similar across iterations, and we verified that output values at these positions are over the predetermined threshold value we mentioned. As a result, it becomes clear that most of the output data can be reused during sparse iterations while some points need to be kept recomputed. Determining these thresholds, which vary across iterations and transformer blocks, does not require additional training. We can determine these local threshold values through empirical experiments and apply them during runtime.

In summary, in addition to the intra-iteration output sparsity achieved through EP (see Section~\ref{section2_2}), the {\agthmsFFN} algorithm introduces a novel inter-iteration output sparsity in diffusion models, providing more opportunities to reduce the computation in FFN layers dramatically. Consequently, by utilizing both output sparsities with similar patterns, computations in any diffusion model comprising transformer blocks can now be reduced without the need for retraining.

\subsection{{\agthmsCD} Mechanism for Data Compaction}
\label{section3_2}

While the number of operations in transformer blocks has been significantly reduced, conventional GPUs cannot efficiently leverage the output sparsity. Since GPUs can utilize structured sparsity well, they cannot translate the unstructured inter- and intra-iteration sparsity into high performance and energy efficiency. Additionally, prior works~\cite{qin2020sigma, yangtrapezoid} have addressed fine-grained sparsity in either input or weight data by placing additional networks between the memory and computation engines. Although these methods are efficient, the sparsity in diffusion models is different since it only exists in the output data, while input and weight data remain dense. Therefore, many existing mechanisms and dedicated accelerators~\cite{qin2020sigma, yangtrapezoid, zhang2020sparch, pal2018outerspace} designed for input or weight sparsity are not suitable for efficiently handling output sparsity. To solve this problem, we introduce the data compaction mechanism, {\agthmsCD}, which consists of condensing and merging.

\begin{figure}[t]
\centering
\includegraphics[width=3.4in]{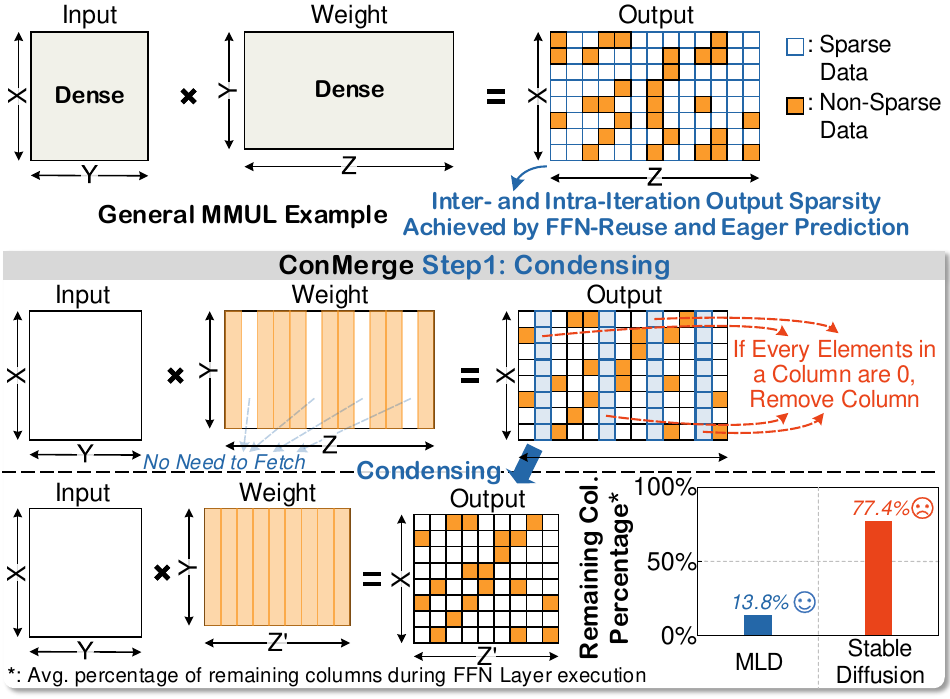}
\caption{Process of Condensing}
\label{fig8}
\vspace{-0.1in}
\end{figure}

\textbf{Condensing.}
Firstly, we propose a simple but efficient strategy named condensing. As shown in Figure~\ref{fig8}, when all elements in a column are sparse, the condensing process removes the corresponding column. This reduces the number of required operations in the MMUL proportionally to the reduction in columns. Moreover, it decreases the required external memory accesses for fetching weight data. The graph in the figure shows the percentage of remaining columns after condensing in MLD~\cite{chen2023executing} and Stable Diffusion models. In the case of MLD, condensing is highly efficient, significantly reducing the number of columns and leaving only 13.8\% of the columns.

However, condensing alone can be less effective. For instance, compared to MLD, Stable Diffusion has a relatively large number of rows in the output matrix, making it less likely for all the elements in a column to be zero, even though the overall output sparsity is high. Consequently, the remaining column percentage after condensing is still 77.4\%, indicating the need for an additional method.

\textbf{Merging.}
To efficiently handle the remaining output sparsity after condensing, we propose a method called merging. Before delving into merging, it is essential to consider the MMUL from an HW perspective. Typically, an HW accelerator running MMUL operations cannot generate all the output data simultaneously and, therefore, employs a tiling strategy. It involves partitioning the output matrix into multiple tiled blocks, which are executed sequentially.

\begin{figure}[t]
\centering
\includegraphics[width=3.4in]{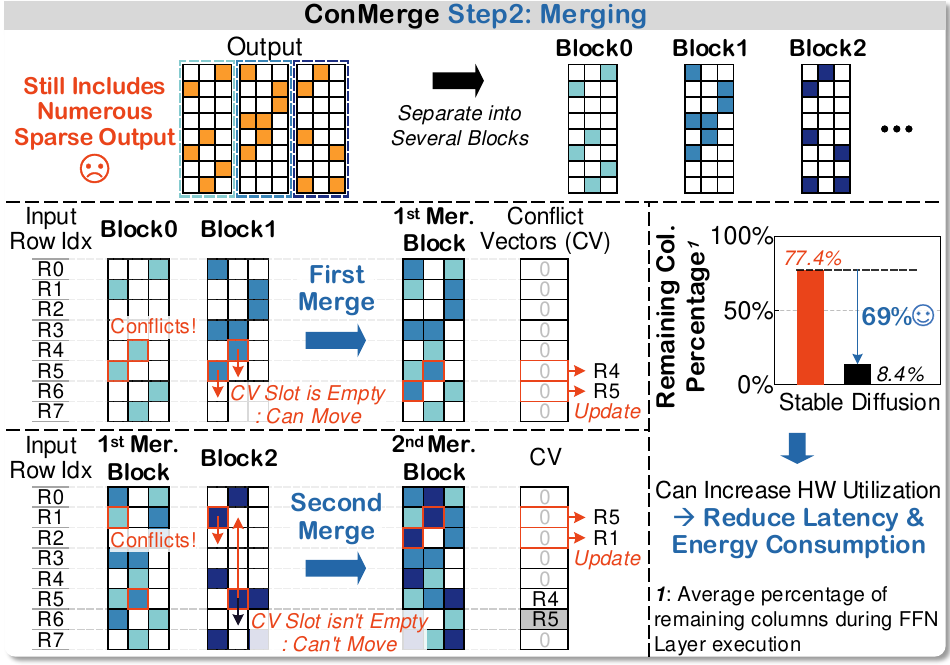}
\caption{Illustration of Merging}
\label{fig9}
\vspace{-0.1in}
\end{figure}

Considering this tiling concept, as illustrated in Figure~\ref{fig9}, the first step of the merging is to partition the output matrix into multiple blocks tailored to the HW configuration. For example, in the toy HW model shown, the merging separates nine columns into three blocks, with each block having a width of three columns. In the second step, we merge two blocks to create a new merged block, thereby reducing the number of tiled blocks that the HW needs to execute. For example, as shown in the figure, we merge Block0 and Block1 to generate the first merged block. Since both blocks share the same row of the input matrix, merging is straightforward on the row side. However, on the column side, the merged block’s columns can originate from either the original weight column of Block0 or Block1. Consequently, every element in the new merged matrix requires tracking its own weight column origin. To manage this efficiently, we incorporate specialized HW components to generate control signals with minimal overhead, with further details provided in Section~\ref{section4_3}.

While merging, conflicts might occur at specific positions when two blocks are not sparse at that position. To solve this problem, we move the elements of the merging block where conflicts occur to other sparse rows within the same column. For instance, if conflicts occur at input rows with indices four (R4) and five (R5), we move the elements of Block1 where conflicts occur to other sparse rows within the same columns, R5 and R6, respectively. Finally, we prepare for the next merging step by updating this information in the conflict vectors (CVs), indicating that in addition to the original input rows R5 and R6, additional input rows R4 and R5 are now required for those rows.

As the last step, we merge the result of the first merged block with the remaining Block2, further reducing the number of tiled blocks. The scenario in the figure's example highlights the necessity of CV. During the second merging phase, conflicts also occur in R1 and R5, but unlike the first merging phase, where all the CV slots are empty, the seventh CV slot is already occupied by R5. Therefore, the mechanism finds another candidate, the R1 row, and finally updates the CV's second slot value as R5.

Although the example illustrates merging with only three blocks, in real scenarios, there are many blocks. If the first merge attempt between Block0 and Block1 fails, the process of merging with Block0 continues with the subsequent blocks, from Block2 to the last block. Once two blocks are successfully merged or all blocks are attempted, the merging moves to the second phase. In the second phase, the merge begins with the first merge result rather than starting from Block0. By generating as many merged blocks as possible, the merging process ultimately succeeds in reducing the remaining columns. For instance, in the Stable Diffusion case, which could not benefit from condensing alone, the percentage of remaining columns dramatically reduced from 77.4\% to 8.4\%, verifying the efficiency of the merging process.

Consequently, by combining the condensing and merging ideas, the {\agthmsCD} mechanism can convert large and sparse output matrices into small and compact matrices, enabling the proposed HW architecture to handle output sparsity with high utilization, thereby reducing latency and total energy consumption. The detailed implementation of the HW architecture and how it addresses the overheads of {\agthmsCD}, such as required cycles and HW resource usage, will be described in the following Section.
\section{{\Arch}'s Hardware Architecture}
\label{section4}

\subsection{Architecture Overview}
\label{section4_1}

\begin{figure}[t]
\centering
\includegraphics[width=3.4in]{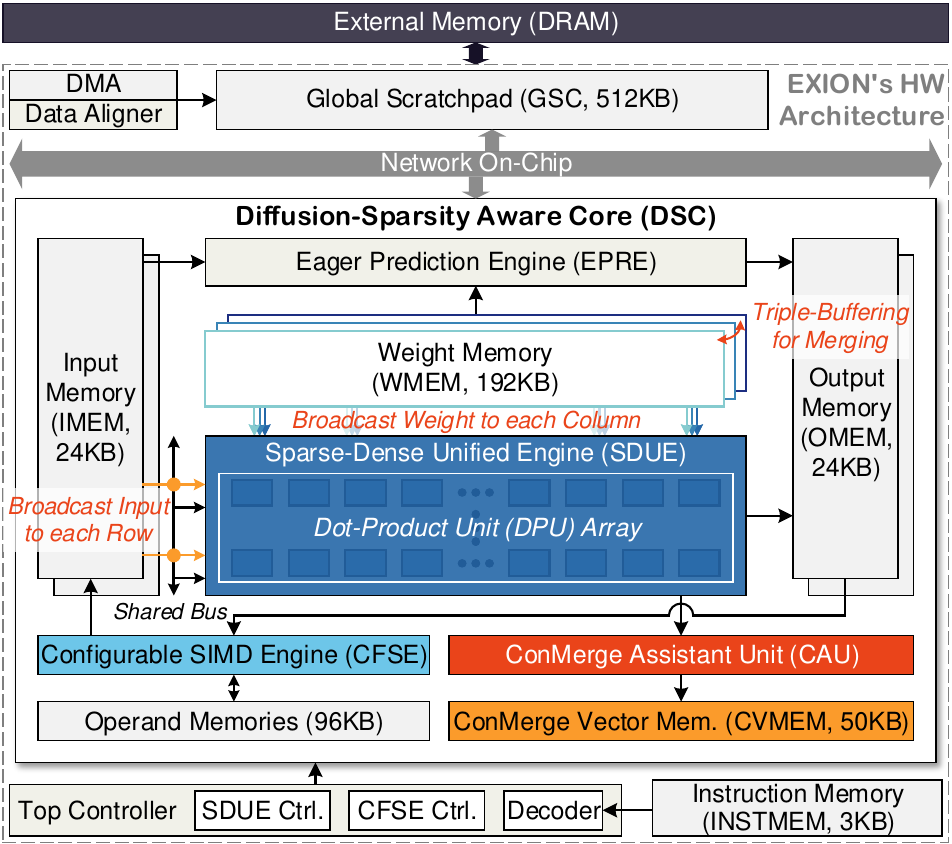}
\caption{Hardware Architecture of {\Arch}}
\label{fig10}
\vspace{-0.1in}
\end{figure}

Figure~\ref{fig10} illustrates the overall architecture that comprises a top controller with instruction memory (INSTMEM), a diffusion-sparsity aware core (DSC), a global scratchpad (GSC) connected to the DSC via a network-on-chip, and a direct memory access (DMA) that stores/loads data to/from external DRAM.

The heart of the architecture is the DSC, which can run the diffusion model's dense and sparse iterations with an identical core. It is equipped with a sparse-dense unified engine (SDUE) comprising a dot-product unit (DPU) array. SDUE can compute the normal dense output matrix and also handle the merged block, which is the result of {\agthmsCD}. To this end, input and weight memories (IMEM and WMEM) are double-buffered and triple-buffered, respectively. This buffering scheme is utilized not only to hide the latency of data fetching but also to broadcast the required data to SDUE. Additionally, the DSC includes a {\agthmsCD} assistant unit (CAU) with {\agthmsCD} vector memory (CVMEM) to generate the control signals. The DSC also includes a configurable SIMD engine (CFSE) with operand memories for accurate computation of special functions such as layer normalization, Softmax, non-linear functions, and residual addition. We design the arithmetic units (ALUs) in CFSE to be configurable, either one-way 32-bit or two-way 16-bit for double throughput. Finally, an eager prediction engine (EPRE) is integrated into DSC to predict the attention score with enhanced accuracy compared to the original EP method.

Operation flow begins by fetching instructions, input, and weight data from the external DRAM to the GSC. Then, the top controller fetches instructions from INSTMEM and, depending on the tiling strategy, unicasts or broadcasts the input and weight to the IMEM and WMEM, respectively. Once the data is ready, the top controller initiates the diffusion process, in which SDUE, EPRE, and CFSE start their computations. During the process, EPRE's latency is mostly hidden by SDUE and CFSE execution due to pipelining schemes. To facilitate this, we adopt an appropriate data mapping and tiling strategy. Moreover, data such as weights and intermediate results are continuously transferred among the DSC, GSC, and external DRAM until the execution is complete.

\subsection{Sparse-Dense Unified Engine}
\label{section4_2}

\begin{figure}[t]
\centering
\includegraphics[width=3.4in]{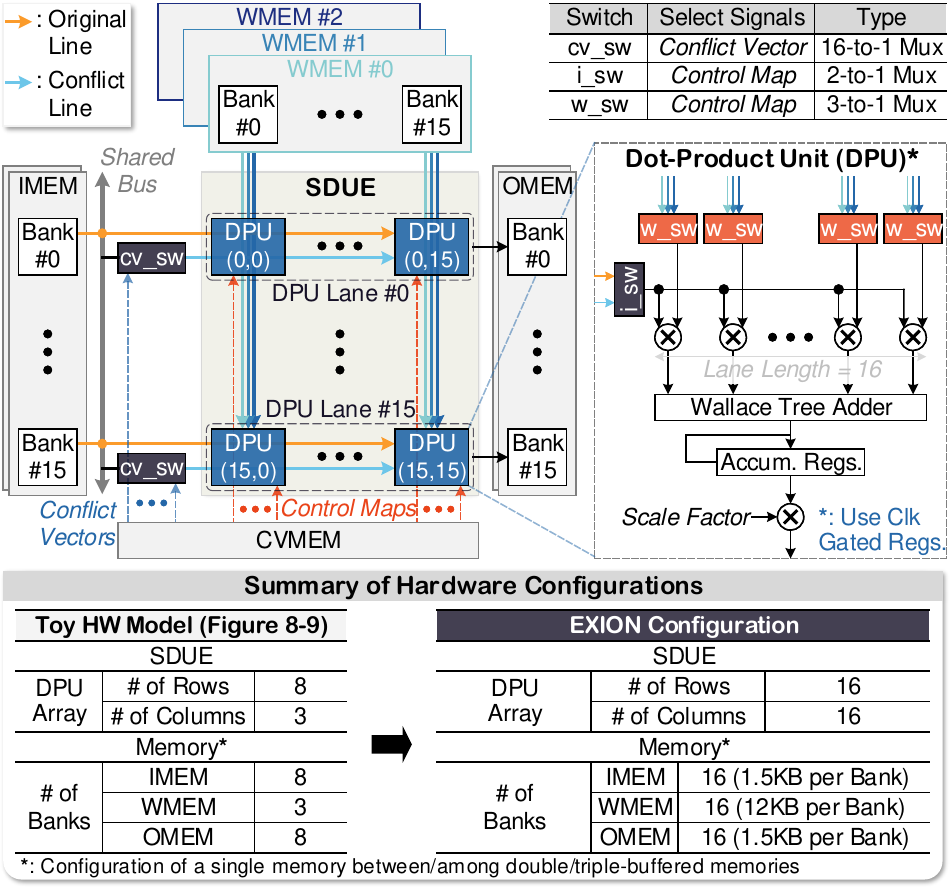}
\caption{Design of Sparse-Dense Unified Engine}
\label{fig11}
\vspace{-0.1in}
\end{figure}

Figure~\ref{fig11} illustrates the design of SDUE, which efficiently computes the most compute-intensive MMUL operations in the transformer block, regardless of the presence of output sparsity. It consists of an array of DPUs, and each DPU contains integer multipliers, followed by Wallace tree adders~\cite{wallace1964suggestion} and accumulation registers. For normal (dense) MMUL operations without any output sparsity, each DPU receives input and weight data from the corresponding memory banks (e.g., DPU$_{(0,0)}$ receives input from IMEM bank \#0 and WMEM bank \#0), executes MAC operations repeatedly and generates the final output of the dot product.

As for the MMUL computation with output sparsity, SDUE maps perfectly to the result of the {\agthmsCD} mechanism. In order to achieve this, each row of the DPU array, named DPU lane, is equipped with a conflict vector switch (cv\_sw). Consequently, each DPU has two input lines broadcasted from two different IMEM banks. We refer to these lines as the original line and the conflict line, based on the originating IMEM bank being broadcasted. For example, all DPUs within DPU lane \#0 directly receive data from IMEM bank \#0 via the original line, while the conflict line data can come from any IMEM bank ranging from \#0 to \#15, as determined by the CV that controls the cv\_sw.

Additionally, each DPU is equipped with two switches: a weight switch (w\_sw) and an input switch (i\_sw). The w\_sw allows the DPU to select among three weight lines broadcasted vertically, each corresponding to three WMEMs, WMEM \#0 to WMEM \#2. Since our {\agthmsCD} mechanism allows merging twice, all the WMEMs, which are triple-buffered, are utilized. Weight data in WMEM's bank corresponds to the column of the weight matrix. Therefore, three columns of weight data, each from the bank with the same index in three WMEMs, are broadcasted to each column of the DPU array (e.g., DPU$_{(0,0)}$ receives weight data from bank \#0 of the three WMEMs). On the other hand, the i\_sw enables the DPU to choose between the two input lines being broadcasted (original or conflict line). Both switches are controlled by control maps (CMs) generated by the CAU, and with a combination of them, each DPU can flexibly compute any combination of dot product operations with respect to the input and weight being broadcasted.

Consequently, the SDUE enables the realization of the {\agthmsCD} mechanism without using large and complex HW resources. This design allows EXION to achieve high utilization and reduce latency. Additionally, clock gating is applied to all the registers in the SDUE's datapath. This strategy addresses any remaining output sparsity after merging, contributing to increased energy efficiency. It is important to note that various accelerators designed to handle fine-grained sparsity in input and weight data often include specialized but complex networks to distribute data to processing engines, which incurs additional latency or area overheads. In contrast, the SDUE broadcasts input and weight data to the DPU array, efficiently supporting the {\agthmsCD} mechanism for fine-grained output sparsity handling with relatively small overhead.

\subsection{{\agthmsCD} Assistant Unit}
\label{section4_3}

\begin{figure}[t]
\centering
\includegraphics[width=3.4in]{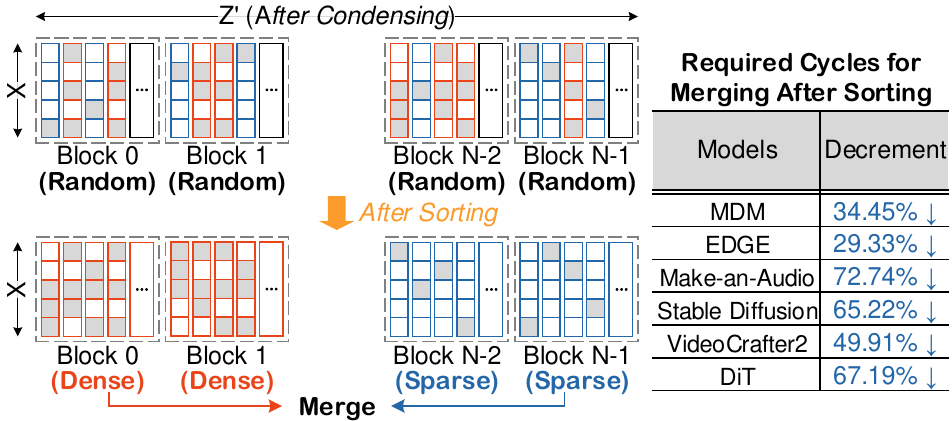}
\caption{Sorting Strategy for Faster Merging}
\label{fig12}
\end{figure}

In the proposed architecture, the CAU generates the {\agthmsCD} vectors that enable the SDUE to address merged blocks by the {\agthmsCD} mechanism. To reduce the energy and latency overhead of these additional generation processes, we execute additional sorting before generating the {\agthmsCD} vectors. As shown in Figure~\ref{fig12}, merging blocks after sorting the columns by the level of sparsity can reduce the total cycles required for the merging process. Compared to merging two blocks with random sparsity, merging two blocks with consideration of their sparsity levels, such as one dense and one sparse, reduces the chances of failure and the need to try merging with other blocks. The figure indicates that this strategy can successfully reduce cycle counts for {\agthmsCD} vector generation, ranging from 29.3\% to 72.7\%. This reduction contributes to lower additional overheads of the {\agthmsCD} mechanism. However, implementing naive sorting can incur additional overheads, such as high latency and area. Therefore, we optimize the buffer in the CAU for efficient sorting.

\begin{figure}[t]
\centering
\includegraphics[width=3.4in]{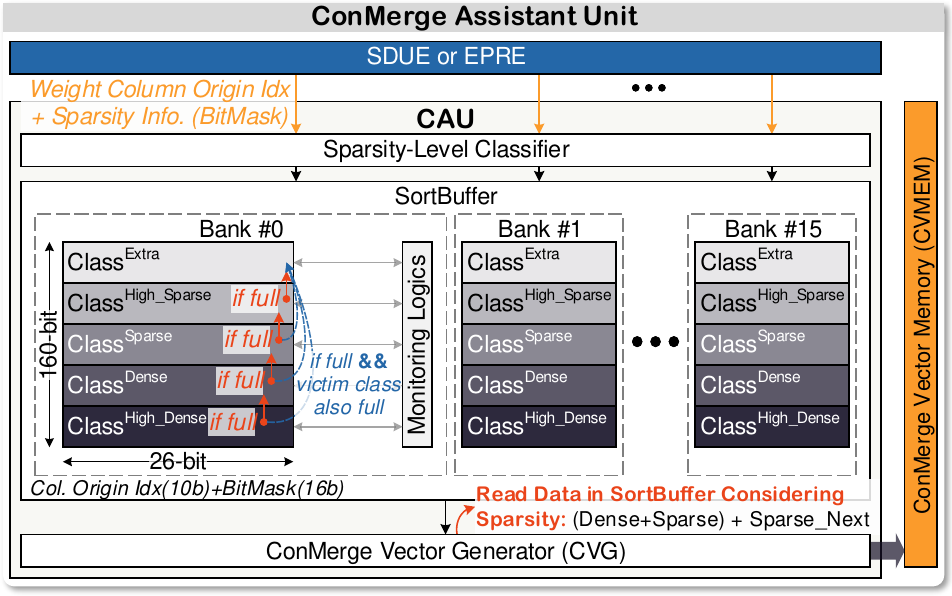}
\caption{Efficient Sorting in CAU}
\label{fig13}
\vspace{-0.1in}
\end{figure}

Figure~\ref{fig13} illustrates the efficient sorting method in the CAU. While SDUE computes the FFN layer of dense iterations, the CAU receives the output matrix's original column index and sparsity information represented as a bitmask from each DPU lane. Then, a sparsity-level classifier first counts the number of non-zero bits in the bitmask and decides the sparsity level of each input data, from high\_dense to high\_sparse. Next, the SortBuffer selects a class and stores the data in the corresponding class based on its sparsity level. During this process, if a class is full, it sends the input bitmask with the column index to the next sparse class, and if that is also full, it sends the bitmask to the extra class. As a result, the bitmask is sorted depending on the sparsity level, not completely but in a coarse manner, which is sufficient to increase the success ratio of merging. Additionally, when data in bitmasks are all zero, those inputs are not stored in the SortBuffer, constituting the condensing in the {\agthmsCD} mechanism.

\begin{figure}[t]
\centering
\includegraphics[width=3.4in]{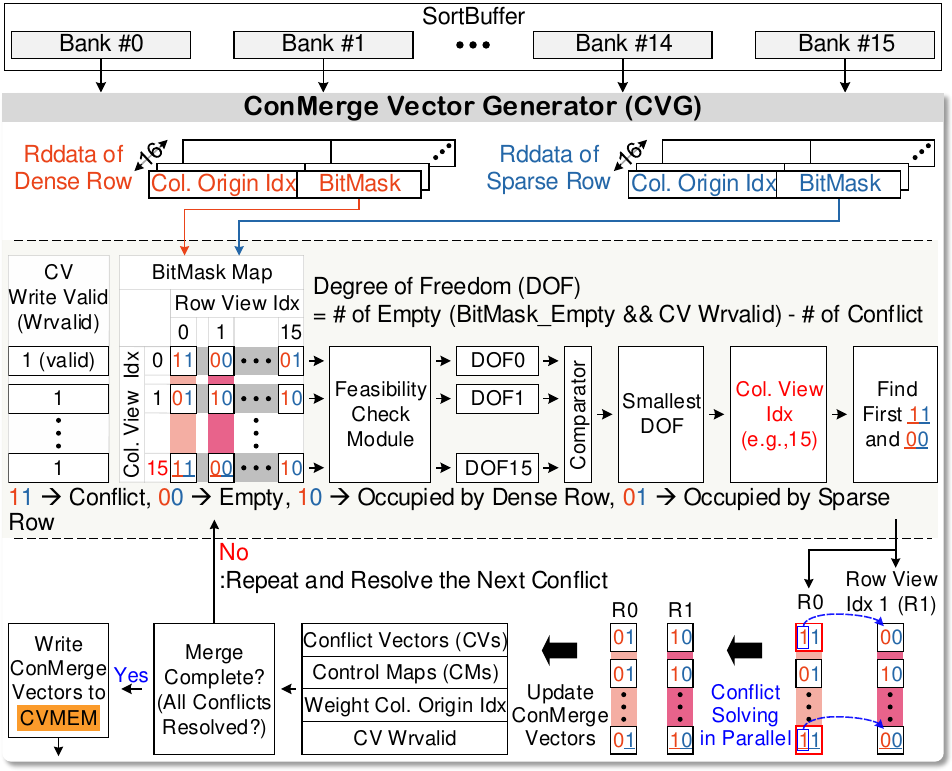}
\caption{Detailed Merging Process in {\agthmsCD} Vector Generator}
\label{fig14}
\vspace{-0.1in}
\end{figure}

Figure~\ref{fig14} shows how {\Arch} generates control signals to support the merging process within the {\agthmsCD} vector generator (CVG). Once sorting is complete, the CVG reads the information in the SortBuffer, selecting one row from the most dense and the other from the most sparse. It then integrates the bitmasks from both rows to create a bitmask map. Using this bitmask map, it identifies problematic columns by calculating the degree of freedom (DOF) for each column view index and selecting the one with the smallest DOF. Within that column, the CVG selects two row view indices by finding the first conflict slot and the first empty slot. Then, it resolves conflicts by moving conflicting slots to empty ones in parallel and subsequently updating control signals such as CVs, CMs, and the weight column origin index. These steps are repeated until all conflicts are resolved. After successfully merging two rows, the CVG continues by merging an additional sparse row from the SortBuffer, writing the final output to CVMEM, and proceeding to the next merging operation. Consequently, the efficient sorting method and subsequent control signal generation in the CAU enable {\Arch} to effectively manage the {\agthmsCD} mechanism overheads while keeping the CAU minimal in size, accounting for only 0.94\% of the DSC area.

\subsection{Eager Prediction Engine}
\label{section4_4}

\begin{figure}[t]
\centering
\includegraphics[width=3.4in]{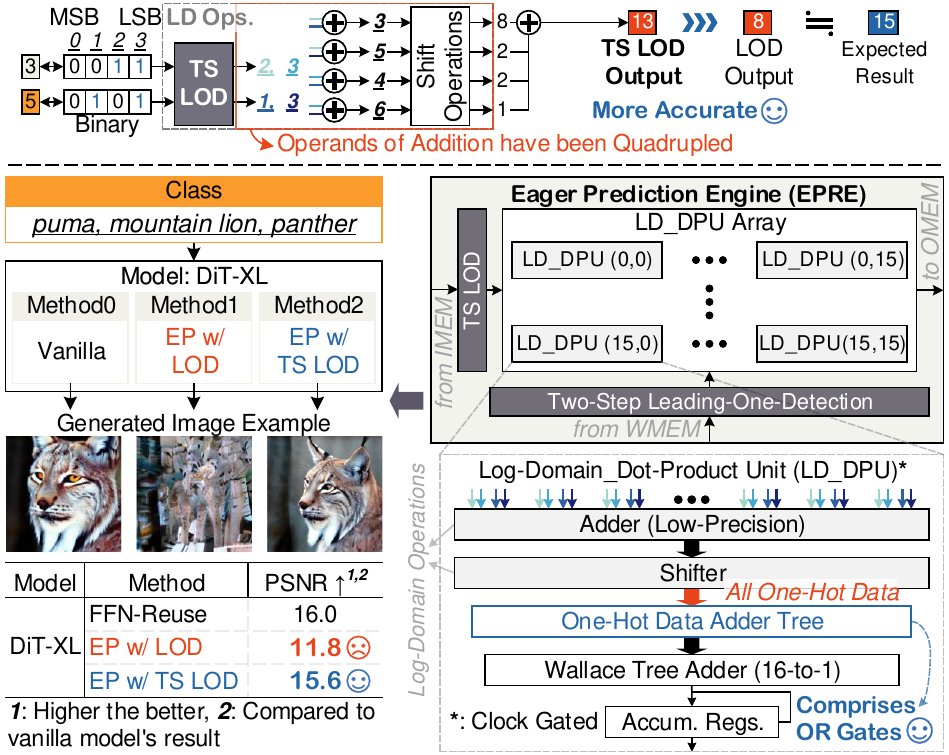}
\caption{EPRE with Two-Step Leading-One Detection}
\label{fig15}
\vspace{-0.1in}
\end{figure}

As aforementioned in Section~\ref{section2}, the SOTA transformer accelerator has proposed the EP method that predicts the attention score with a simple log-domain algorithm. However, directly applying this scheme to diffusion models incurs critical accuracy loss. Figure~\ref{fig15} illustrates image generation with the DiT, showing that EP's leading-one detection (LOD) produces a poor-quality image and incurs a PSNR drop compared to applying the {\agthmsFFN} only.

To overcome this problem, we execute two-step leading-one detection (TS LOD) in EPRE before sending data to the log-domain DPU (LD\_DPU) array. This process first conducts LOD and then detects an additional bit after converting the leading-one bit to zero, finding the positions of two bits in total. We have verified that the PSNR increases from 11.8 to 15.6 with this simple improvement. However, since the operands of addition are quadrupled, the overhead of addition might become a problem with an unoptimized implementation. We exploit the data characteristic in the log domain, where the output of the shift operations is always one-hot data. Consequently, we develop a one-hot data adder tree comprising only simple OR gates in each LD\_DPU, successfully reducing the overhead of the TS LOD method.
\section{Evaluation}
\label{section5}

\newcommand{\Base}{\texttt{{\Arch}\textsuperscript{x}\textunderscore{Base}}}
\newcommand{\EP}{\texttt{{\Arch}\textsuperscript{x}\textunderscore{EP}}}
\newcommand{\FFNR}{\texttt{{\Arch}\textsuperscript{x}\textunderscore{FFNR}}}
\newcommand{\All}{\texttt{{\Arch}\textsuperscript{x}\textunderscore{All}}}

\newcommand{\BaseS}{\texttt{{\Arch}\textsuperscript{4}\textunderscore{Base}}}
\newcommand{\EPS}{\texttt{{\Arch}\textsuperscript{4}\textunderscore{EP}}}
\newcommand{\FFNRS}{\texttt{{\Arch}\textsuperscript{4}\textunderscore{FFNR}}}
\newcommand{\AllS}{\texttt{{\Arch}\textsuperscript{4}\textunderscore{All}}}

\newcommand{\BaseL}{\texttt{{\Arch}\textsuperscript{24}\textunderscore{Base}}}
\newcommand{\EPL}{\texttt{{\Arch}\textsuperscript{24}\textunderscore{EP}}}
\newcommand{\FFNRL}{\texttt{{\Arch}\textsuperscript{24}\textunderscore{FFNR}}}
\newcommand{\AllL}{\texttt{{\Arch}\textsuperscript{24}\textunderscore{All}}}

\newcommand{\AllNew}{\texttt{{\Arch}\textsuperscript{42}\textunderscore{All}}}

\subsection{Accuracy Evaluation}
\label{section5_1}

\textbf{Workloads.}
To demonstrate the flexibility and accuracy of {\Arch}, we selected seven different diffusion models as our workloads, as they encompass all three types of diffusion models (see Figure~\ref{fig3}). These models include various applications such as text-to-motion (MLD~\cite{chen2023executing} and MDM~\cite{tevet2023human}), music-to-motion (EDGE~\cite{tseng2023edge}), text-to-image (Stable Diffusion~\cite{rombach2022high}), class-to-image (DiT~\cite{peebles2023scalable}), text-to-audio (Make-an-Audio~\cite{huang2023make}), and text-to-video (VideoCrafter2~\cite{chen2024videocrafter2}) generations.

\textbf{Accuracy.}
Since the rationale for achieving output sparsity in diffusion models is fundamentally based on an approximation strategy, we focused on verifying that the accuracy loss incurred by the proposed {\Arch} is minimal. To this end, we tested the model accuracy in various diffusion models with the given dataset. Additionally, to demonstrate that the proposed solution in {\Arch} can be generally adapted to any diffusion model during inference, we did not conduct any retraining. We also determined values such as the threshold in {\agthmsFFN} and the top-k selection ratio through empirical experiments.

\begin{table*}[t]
\caption{Model Accuracy Evaluation}
\centering
\includegraphics[width=7.0in]{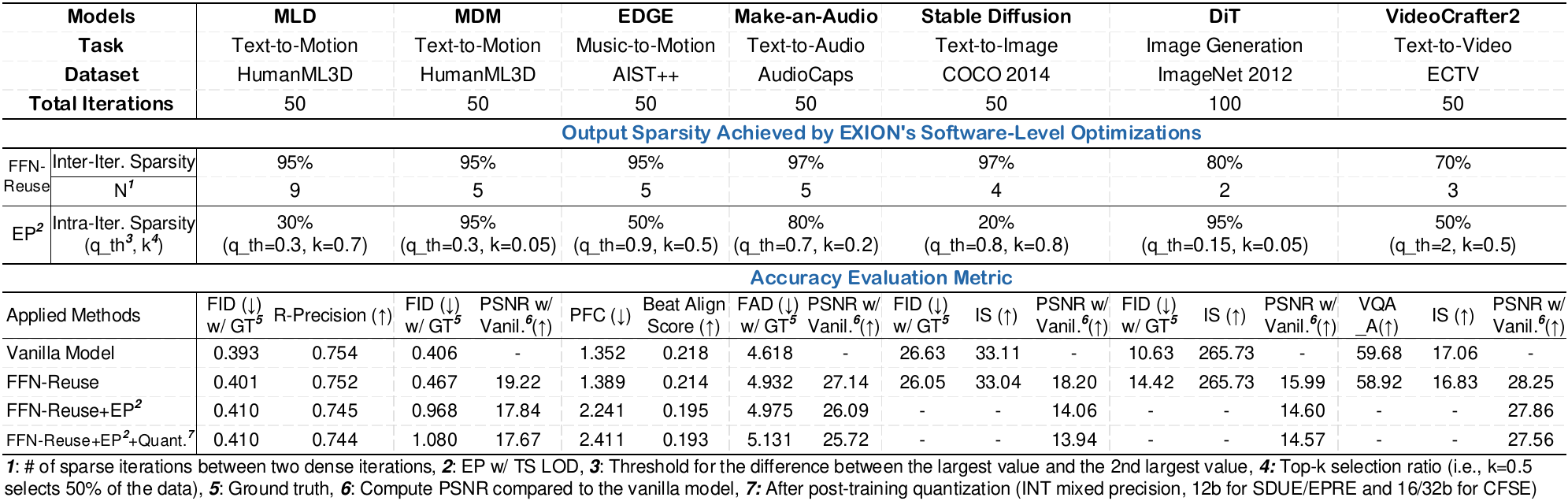}
\label{table1}
\vspace{-0.1in}
\end{table*}

\begin{figure}[t]
\centering
\includegraphics[width=3.4in]{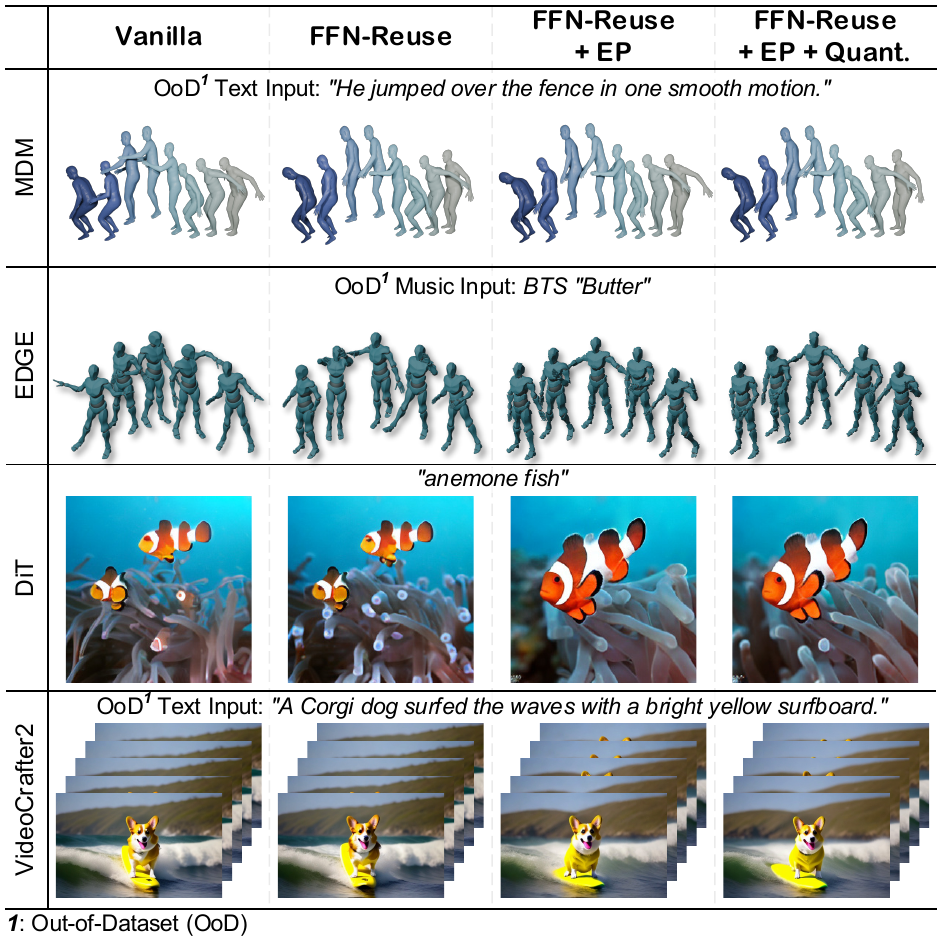}
\caption{Visualization of Generated Outputs}
\label{fig16}
\vspace{-0.1in}
\end{figure}

Table~\ref{table1} shows the results of the accuracy evaluation experiments. They indicate that the proposed {\agthmsFFN} algorithm, which achieves inter-iteration sparsity of 70-97\%, has no impact on accuracy. Across all the benchmarks, the evaluation metrics in each model show a minimal difference compared to the results of the vanilla model. This ensures that the {\agthmsFFN} is a feasible solution and can be applied to other diffusion models beyond our benchmarks.

In addition to {\agthmsFFN}, adopting the EP method achieves additional intra-iteration sparsity of 20-95\%. This approach also shows trivial differences compared to the results of the vanilla model. However, there are exceptions for both MDM and EDGE models. For these two models, one of the two different metrics shows a trivial difference while the other does not. Despite this, Figure~\ref{fig16}, which visualizes the generated outputs, demonstrates that the quality of the results remains uncompromised even when one metric indicates poorer accuracy. Moreover, these experiments were conducted using out-of-dataset inputs, except for the DiT model, indicating that {\Arch}'s optimizations are applicable to production environments. Additionally, we verified that accuracy is maintained after applying post-training quantization, reducing MMUL operations to 12-bit INT and other operations to either 16-bit or 32-bit INT, aligning with our HW architecture.

\subsection{Experimental Setup}
\label{section5_2}

\begin{table}[t]
    \caption{Hardware Specifications of GPUs and EXION}
    \centering
    \includegraphics[width=3.4in]{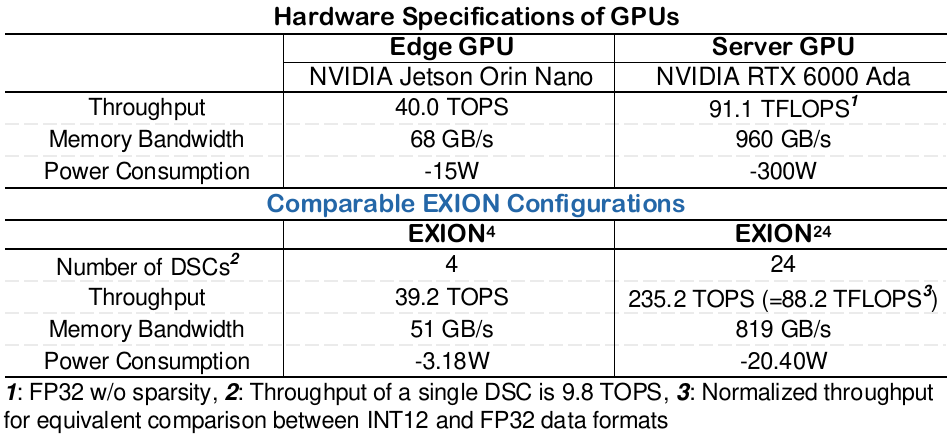}
    \label{table2}
    \vspace{-0.1in}
\end{table}

\textbf{Performance.}
To evaluate the performance of the {\Arch} accelerator, we developed a custom cycle-level simulator and integrated it with Ramulator~\cite{kim2015ramulator}, a widely used DRAM simulator, to model DRAM latency. For more comprehensive results, we evaluated the performance of {\Arch} against two different settings: the edge GPU (NVIDIA Jetson Orin Nano~\cite{jetson_orin_nano}) and the server GPU (NVIDIA RTX 6000 Ada~\cite{rtx6000}). To ensure a fair comparison, {\Arch} was instantiated in two distinct configurations, summarized in Table~\ref{table2}. To match the peak performance and DRAM bandwidth of the edge GPU, which are 40TOPS and 68GB/s using LPDDR5, we instantiated \texttt{{\Arch}\textsuperscript{4}} with 4DSCs and 51GB/s DRAM bandwidth using LPDDR5. For the server GPU, which has 91TFLOPS and 960GB/s using GDDR6, we instantiated \texttt{{\Arch}\textsuperscript{24}} with 24DSCs and 819GB/s DRAM bandwidth.

Also, for the ablation study, we evaluated {\Arch} by applying the proposed methods in this paper incrementally. While the \texttt{x} indicates the number of DSCs, we tested in four different configurations: {\Base} without any optimizations, {\EP} with EP, {\FFNR} with {\agthmsFFN}, and {\All} with both EP and {\agthmsFFN}.

\textbf{Power/Area.}
To estimate the power consumption and area of {\Arch}, we first implemented its design with a single DSC at the RTL level using System Verilog and then synthesized it using Synopsys's Design Compiler~\cite{synopsys_design_complier} with 14nm process technology. We verified that it operates at 0.8V and an 800MHz clock frequency without any timing issues. Values from the synthesized results are used as the power consumption and area usage of {\Arch}. For power modeling of LPDDR5 DRAM, we utilized the energy information provided by the DRAM vendor~\cite{kim2016future, lee2017understanding}. Finally, we integrated these results into our custom simulator, accurately measuring the power consumption and area of {\Arch}. Moreover, to measure the power consumption of GPUs, we utilized the NVIDIA-SMI~\cite{nvidia_smi} and tegrastats~\cite{tegrastats_Utility} tools.

\subsection{Performance Evaluation}
\label{section5_3}

\begin{figure}[t]
\centering
\includegraphics[width=3.4in]{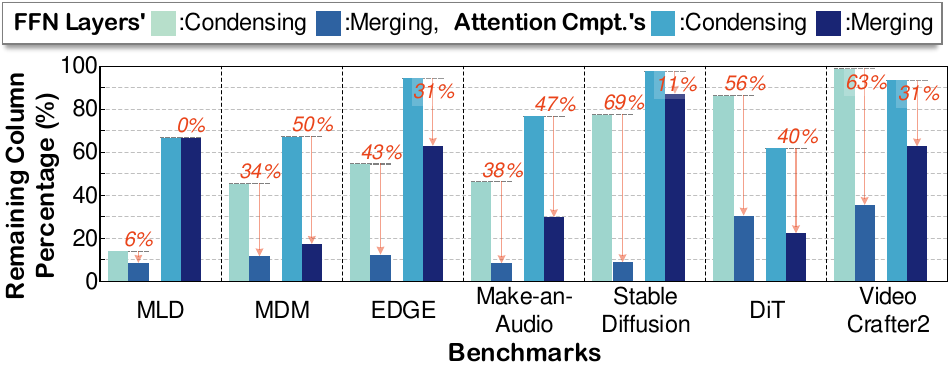}
\caption{{\agthmsCD} Efficiency}
\label{fig17}
\vspace{-0.1in}
\end{figure}

\textbf{{\agthmsCD} Efficiency.}
Figure~\ref{fig17} shows the efficiency of the proposed {\agthmsCD} mechanism, comprising condensing and merging. To verify their efficiency in each step, the figure displays the remaining column percentage in the output matrix of the first FFN layer and the attention score after applying condensing and then merging. For the condensing case, the average percentage of remaining columns in the first FFN layer and attention score is 60.3\% and 80.0\%, respectively. Although small models such as MLD and MDM can significantly reduce columns with condensing alone, the average value appears high. This is because, in models where the row size of the output matrix is large, there is less chance that all elements in a column are sparse. For instance, although the output sparsity is high at 97\% and 70\% in Stable Diffusion and Videocrafter2, respectively, 77.4\% and 98.6\% of output columns in the FFN layer still remain.

Merging solves this issue by significantly reducing these remaining columns. The average percentage of remaining columns in the first FFN layer and attention score achieve 16.2\% and 50.0\%, respectively, after merging. Notably, for the problematic models, only 8.4\% and 35.2\% of columns in the FFN layer are left for Stable Diffusion and VideoCrafter2, respectively. These results verify that the proposed {\agthmsCD} successfully enables the {\Arch} to handle output matrices with high sparsity.

\begin{figure}[t]
\centering
\includegraphics[width=3.4in]{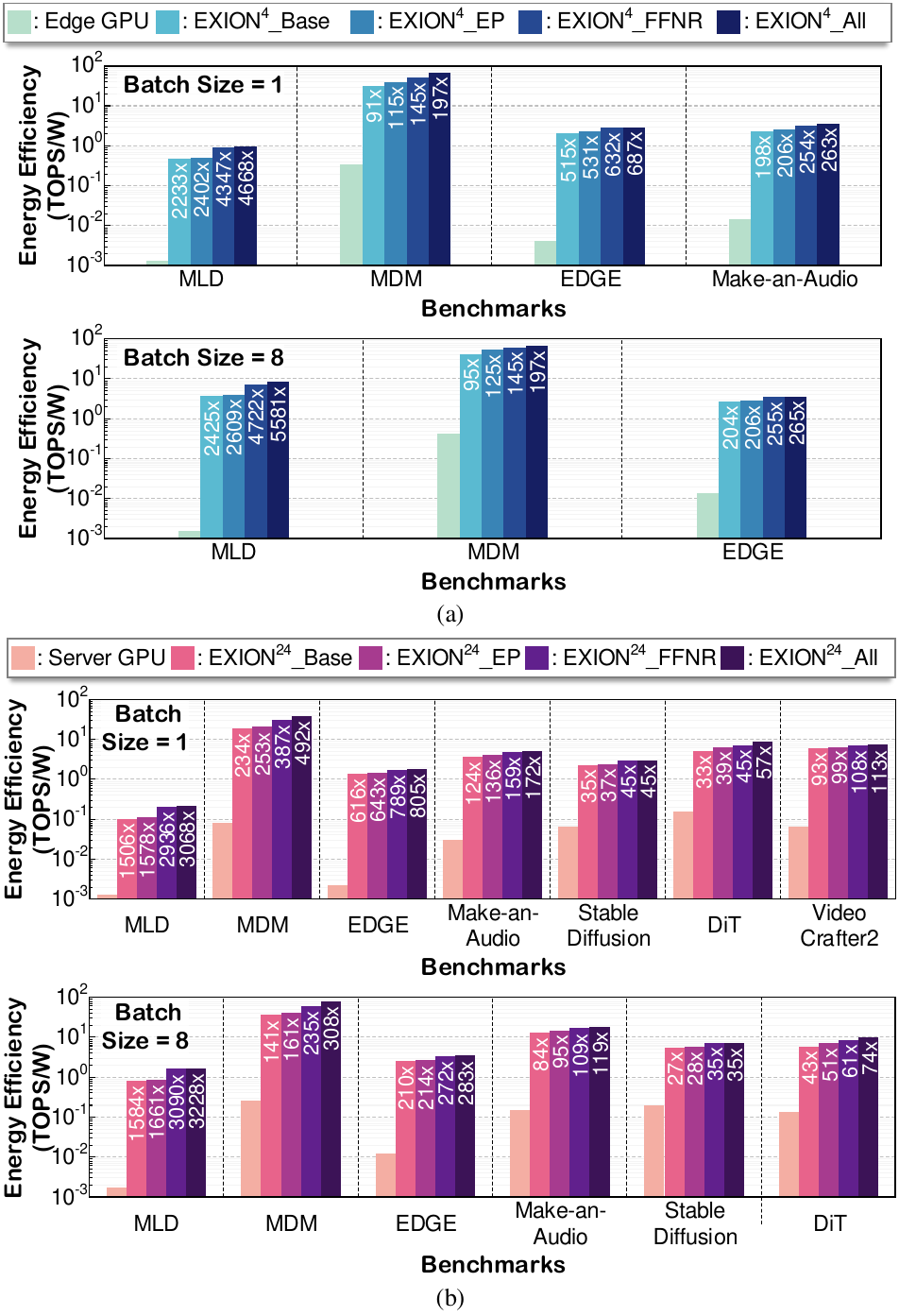}
\caption{Energy Efficiency Comparison versus (a) Edge GPU (b) Server GPU}
\label{fig18}
\vspace{-0.1in}
\end{figure}

\textbf{Energy Efficiency.}
To demonstrate that {\Arch} is an energy-efficient solution, we compared its energy efficiency with GPUs. Specifically, the experiments were conducted using batch sizes of one and eight, and ablation studies were performed to analyze the effect of each optimization method that {\Arch} introduces. Figure~\ref{fig18} (a) shows the comparison results of \texttt{{\Arch}\textsuperscript{4}} versus the edge GPU, where large models are not considered since executing them on an edge GPU is infeasible due to insufficient memory size. The experimental results with batch size one reveal that {\BaseS} achieves 0.5-31.4 TOPS/W energy efficiency within benchmarks, which is 91.1-2232.6$\times$ more energy efficient than the edge GPU. Utilizing the resulting intra-iteration output sparsity by the EP method enables the efficiency gain to increase to 114.7-2401.7$\times$. Notably, leveraging the inter-iteration output sparsity achieved by {\agthmsFFN} excels this gain, increasing it by 145.0-4347.2$\times$. This reveals that optimizing the FFN layers in diffusion models is crucial. Finally, by utilizing both output sparsity types present in diffusion models, {\AllS} achieves 1.0-67.8 TOPS/W energy efficiency, which is 196.9-4668.2$\times$ more energy efficient than the edge GPU.

Figure~\ref{fig18} (b) shows the comparison results of \texttt{{\Arch}\textsuperscript{24}} versus the server GPU. While most of the results show that the energy efficiency of {\Arch} is significantly higher than that of the server GPU, its efficiency gain tends to drop in models such as Make-an-Audio and Stable Diffusion. This reduction is due to the presence of ResBlocks in these models, where we have not utilized any sparsity optimizations. However, considering the recent trend in ML research towards developing transformer-only architectures, the efficiency of {\Arch} is still assured. Finally, {\AllL} achieves 0.2-38.1 TOPS/W energy efficiency with batch size one, which is 45.1-3067.6$\times$ more energy efficient than the server GPU, proving that the scale-out version of the {\Arch} architecture is also energy-efficient and can be a promising solution for large models typically executed in servers. Additionally, the experimental results in Figures~\ref{fig18} (a) and (b) with batch size eight show that {\Arch} is more efficient than the edge and server GPUs regardless of batch size, verifying its effectiveness in various inference settings.

\begin{figure}[t]
\centering
\includegraphics[width=3.4in]{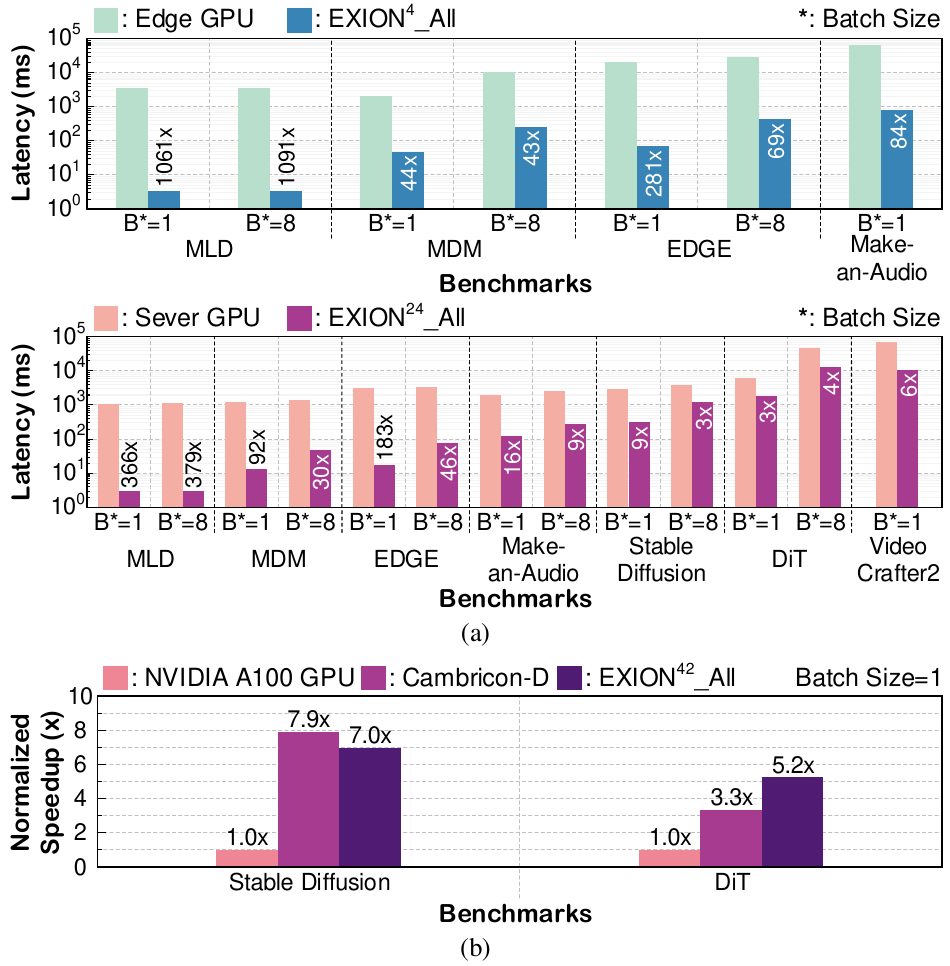}
\caption{(a) Latency Comparison versus Edge GPU and Server GPU and (b) Comparison of Speedup Over Different Accelerators}
\label{fig19}
\vspace{-0.1in}
\end{figure}

\textbf{Latency.}
Figure~\ref{fig19} (a) shows that the proposed {\Arch} successfully reduces the long latency of diffusion models by utilizing inter- and intra-iteration output sparsity. The experiment compares the latency of {\AllS} and {\AllL} with the edge GPU and server GPU, respectively, using two different batch sizes: one and eight. With batch size one, the speedup of {\AllS} and {\AllL} achieves 43.7-1060.6$\times$ and 3.3-365.6$\times$, respectively. Notably, even with batch size eight, they achieve 42.6-1090.9$\times$ and 3.2-379.3$\times$ speedup over GPUs, respectively. This demonstrates that utilizing output sparsity through the {\agthmsCD} mechanism can dramatically reduce execution time in various settings.

\textbf{Comparison with SOTA Accelerator.}
Figure~\ref{fig19} (b) compares the speedup of {\Arch} and Cambricon-D~\cite{kong2024cambricon}, a SOTA diffusion accelerator, over the NVIDIA A100 GPU~\cite{nvidia_a100}. Cambricon-D, specialized for accelerating image generation tasks, uses differential acceleration~\cite{mahmoud2018diffy} on convolutional layers with a systolic array-based design to improve latency and energy efficiency through optimized memory access. We compare {\AllNew}, with 42 DSCs and a memory bandwidth of 1935 GB/s, to Cambricon-D, as both have similar throughput and memory bandwidth. For Stable Diffusion, which includes convolutional layers, Cambricon-D achieves a speedup of 7.9$\times$, slightly exceeding {\AllNew}'s speedup of 7.0$\times$. However, for the DiT model, which lacks convolutional layers in the denoising iterations, {\AllNew} achieves a speedup of 5.2$\times$, outperforming Cambricon-D's 3.3$\times$ speedup. This result highlights our solution's ability to exploit output sparsity within transformer blocks.

It's worth noting that most SOTA image generation models~\cite{peebles2023scalable, chen2023pixart, zheng2023fasttrain} are transitioning from networks including ResBlocks towards transformer-block architectures, which limits the applicability of Cambricon-D's acceleration. Additionally, the optimizations in {\Arch} can be further applied to other tasks, such as text- or audio-to-motion generations, where transformer blocks dominate the operations, positioning {\Arch} as a versatile diffusion accelerator.

\subsection{Power and Area Breakdown}
\label{section5_4}

\begin{table}[t]
\caption{Breakdown of Power and Area Usage}
\centering
\includegraphics[width=3.4in]{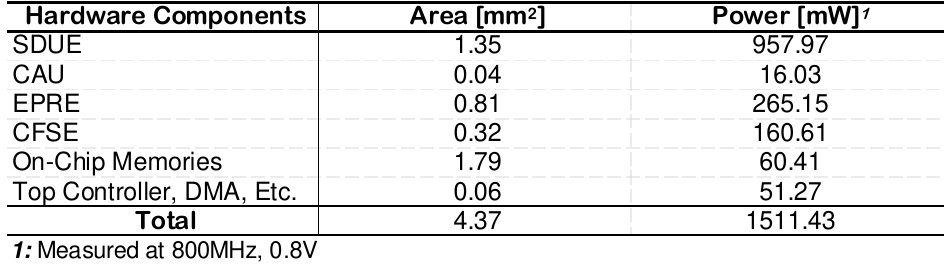}
\label{table3}
\vspace{-0.1in}
\end{table}

Table~\ref{table3} breaks down the power consumption and area of the {\Arch} equipped with a single DSC. Comparison with the server GPU verifies that the architecture of {\Arch} is area-efficient. Specifically, the area of \texttt{{\Arch}\textsuperscript{24}}, which contains 24 DSCs and 64MB GSC, occupies 152.28mm\(^2\). This is relatively smaller compared to the server GPU with a die area of 609mm\(^2\)~\cite{rtx6000area}. In terms of power consumption, {\Arch} with a single DSC consumes 1511.43mW in total. Considering the peak power consumption of the server GPU and edge GPU are 300W and 15W, respectively, the power consumption of {\Arch} is significantly lower even when scaled out. Furthermore, given the dramatic reduction in energy consumption thanks to utilizing output sparsity, it is acceptable that additional HW components such as EPRE and CAU for sparsity handling consume up to 18.6\% of the total power.
\section{Related Works}
\label{section6}

\textbf{Software-Based Approaches.}
A few SW-based approaches~\cite{salimans2022progressive, chung2022come, lu2022dpm, song2023consistency} propose fast sampling methods that reduce the number of iterations. However, without retraining, the reduction is limited in achieving acceptable sampling quality~\cite{zheng2023fast}. $\Delta$-DiT~\cite{chen2024delta} suggests skipping computations of certain DiT blocks by reusing cached output data across iterations. Despite these advancements, as these methods rely on GPU-based optimizations, they have not leveraged the opportunities presented by the significant yet unstructured inter- and intra-iteration sparsity. {\Arch} successfully exploits them by proposing the {\agthmsCD} mechanism along with a dedicated HW architecture.

\textbf{Transformer Accelerators.}
Recently, a large body of work~\cite{ham20203, ham2021elsa, qu2022dota, lu2021sanger, wang2022energy} has been proposed to accelerate attention computation in transformer blocks, primarily targeting language models. However, FFN layers, which have the largest number of operations in recent diffusion models, have not yet been adequately addressed. Although methods handling FFN layers like FACT~\cite{qin2023fact}, which reduces energy consumption via mixed-precision, and SpAtten~\cite{wang2021spatten}, which reduces computations via token pruning, can be applied to diffusion models, the high sparsity present in diffusion models is not yet employed. {\Arch} achieves inter-iteration sparsity by reusing output data across iterations and directly translates this high sparsity into improvements in energy efficiency and performance.

\textbf{SpGEMM Accelerators.}
Various generalized sparse matrix-matrix multiplication (SpGEMM) accelerators~\cite{zhang2020sparch, pal2018outerspace, srivastava2020matraptor, zhang2021gamma, hegde2019extensor, li2023spada} have been proposed to exploit sparsity present in MMUL operations. With the optimization of datapaths and hardware for sparsity handling, they can accelerate SpGEMM operations. SIGMA~\cite{qin2020sigma} and Trapezoid~\cite{yangtrapezoid} propose methods to handle both sparse and dense operations by placing additional networks between memory and computational engines. However, the unique output sparsity present in diffusion models has yet to be addressed. {\Arch} solves this problem by introducing the {\agthmsCD} mechanism and a dedicated hardware architecture that efficiently broadcasts both input and weight data.
\section{Conclusion}
\label{section7}

In this work, we present {\Arch}, the first software-hardware co-designed diffusion accelerator that solves the computation challenges of excessive iterations by exploiting the unique inter- and intra-iteration output sparsity in diffusion models. To this end, we propose two software-level optimizations. First, we propose the FFN-Reuse algorithm that identifies and skips redundant computations in FFN layers across different iterations. Second, we use a modified eager prediction method that employs two-step leading-one detection to predict the attention score accurately.
To handle the resulting unstructured sparsity, we introduce the data compaction mechanism,  {\agthmsCD}, to condense and merge sparse matrices into compact forms. {\Arch}'s dedicated architecture supports this, translating high output sparsity into improved energy efficiency and performance. Our evaluation shows that {\Arch} achieves significant improvements in performance and energy efficiency, with gains of up to 379.3$\times$ and 3067.6$\times$ over a server GPU and up to 1090.9$\times$ and 4668.2$\times$ over an edge GPU.

\ifdefined\hpcacameraready
\section*{Acknowledgements}
This work was supported in part by the Ministry of Science and ICT (MSIT), South Korea, under the Institute of Information \& Communications Technology Planning \& Evaluation (IITP) grants (No. 2022-0-01037, Development of High Performance Processing-In-Memory Technology based on DRAM) and (No. IITP-2025-RS-2023-00256472, the Graduate School of Artificial Intelligence Semiconductor), and by Samsung Electronics.
\fi







\begin{thebibliography}{10}
\providecommand{\url}[1]{#1}
\csname url@samestyle\endcsname
\providecommand{\newblock}{\relax}
\providecommand{\bibinfo}[2]{#2}
\providecommand{\BIBentrySTDinterwordspacing}{\spaceskip=0pt\relax}
\providecommand{\BIBentryALTinterwordstretchfactor}{4}
\providecommand{\BIBentryALTinterwordspacing}{\spaceskip=\fontdimen2\font plus
\BIBentryALTinterwordstretchfactor\fontdimen3\font minus \fontdimen4\font\relax}
\providecommand{\BIBforeignlanguage}[2]{{%
\expandafter\ifx\csname l@#1\endcsname\relax
\typeout{** WARNING: IEEEtranS.bst: No hyphenation pattern has been}%
\typeout{** loaded for the language `#1'. Using the pattern for}%
\typeout{** the default language instead.}%
\else
\language=\csname l@#1\endcsname
\fi
#2}}
\providecommand{\BIBdecl}{\relax}
\BIBdecl

\bibitem{videoworldsimulators2024}
\BIBentryALTinterwordspacing
T.~Brooks, B.~Peebles, C.~Holmes, W.~DePue, Y.~Guo, L.~Jing, D.~Schnurr, J.~Taylor, T.~Luhman, E.~Luhman, C.~Ng, R.~Wang, and A.~Ramesh, \emph{Video generation models as world simulators}, 2024. [Online]. Available: \url{https://openai.com/research/video-generation-models-as-world-simulators}
\BIBentrySTDinterwordspacing

\bibitem{chen2024videocrafter2}
H.~Chen, Y.~Zhang, X.~Cun, M.~Xia, X.~Wang, C.~Weng, and Y.~Shan, ``Videocrafter2: Overcoming data limitations for high-quality video diffusion models,'' in \emph{Proceedings of the IEEE/CVF Conference on Computer Vision and Pattern Recognition}, 2024, pp. 7310--7320.

\bibitem{chen2023pixart}
J.~Chen, J.~Yu, C.~Ge, L.~Yao, E.~Xie, Y.~Wu, Z.~Wang, J.~Kwok, P.~Luo, H.~Lu \emph{et~al.}, ``Pixart-$\alpha$: Fast training of diffusion transformer for photorealistic text-to-image synthesis,'' \emph{arXiv preprint arXiv:2310.00426}, 2023.

\bibitem{chen2024delta}
P.~Chen, M.~Shen, P.~Ye, J.~Cao, C.~Tu, C.-S. Bouganis, Y.~Zhao, and T.~Chen, ``$\delta$-dit: A training-free acceleration method tailored for diffusion transformers,'' \emph{arXiv preprint arXiv:2406.01125}, 2024.

\bibitem{chen2023executing}
X.~Chen, B.~Jiang, W.~Liu, Z.~Huang, B.~Fu, T.~Chen, and G.~Yu, ``Executing your commands via motion diffusion in latent space,'' in \emph{Proceedings of the IEEE/CVF Conference on Computer Vision and Pattern Recognition}, 2023, pp. 18\,000--18\,010.

\bibitem{chung2022come}
H.~Chung, B.~Sim, and J.~C. Ye, ``Come-closer-diffuse-faster: Accelerating conditional diffusion models for inverse problems through stochastic contraction,'' in \emph{Proceedings of the IEEE/CVF Conference on Computer Vision and Pattern Recognition}, 2022, pp. 12\,413--12\,422.

\bibitem{elizalde2023clap}
B.~Elizalde, S.~Deshmukh, M.~Al~Ismail, and H.~Wang, ``Clap learning audio concepts from natural language supervision,'' in \emph{ICASSP 2023-2023 IEEE International Conference on Acoustics, Speech and Signal Processing (ICASSP)}.\hskip 1em plus 0.5em minus 0.4em\relax IEEE, 2023, pp. 1--5.

\bibitem{ham20203}
T.~J. Ham, S.~J. Jung, S.~Kim, Y.~H. Oh, Y.~Park, Y.~Song, J.-H. Park, S.~Lee, K.~Park, J.~W. Lee \emph{et~al.}, ``A\^{} 3: Accelerating attention mechanisms in neural networks with approximation,'' in \emph{2020 IEEE International Symposium on High Performance Computer Architecture (HPCA)}.\hskip 1em plus 0.5em minus 0.4em\relax IEEE, 2020, pp. 328--341.

\bibitem{ham2021elsa}
T.~J. Ham, Y.~Lee, S.~H. Seo, S.~Kim, H.~Choi, S.~J. Jung, and J.~W. Lee, ``Elsa: Hardware-software co-design for efficient, lightweight self-attention mechanism in neural networks,'' in \emph{2021 ACM/IEEE 48th Annual International Symposium on Computer Architecture (ISCA)}.\hskip 1em plus 0.5em minus 0.4em\relax IEEE, 2021, pp. 692--705.

\bibitem{hegde2019extensor}
K.~Hegde, H.~Asghari-Moghaddam, M.~Pellauer, N.~Crago, A.~Jaleel, E.~Solomonik, J.~Emer, and C.~W. Fletcher, ``Extensor: An accelerator for sparse tensor algebra,'' in \emph{Proceedings of the 52nd Annual IEEE/ACM International Symposium on Microarchitecture}, 2019, pp. 319--333.

\bibitem{hendrycks2016gaussian}
D.~Hendrycks and K.~Gimpel, ``Gaussian error linear units (gelus),'' \emph{arXiv preprint arXiv:1606.08415}, 2016.

\bibitem{ho2020denoising}
J.~Ho, A.~Jain, and P.~Abbeel, ``Denoising diffusion probabilistic models,'' \emph{Advances in neural information processing systems}, vol.~33, pp. 6840--6851, 2020.

\bibitem{huang2023make}
R.~Huang, J.~Huang, D.~Yang, Y.~Ren, L.~Liu, M.~Li, Z.~Ye, J.~Liu, X.~Yin, and Z.~Zhao, ``Make-an-audio: Text-to-audio generation with prompt-enhanced diffusion models,'' in \emph{International Conference on Machine Learning}.\hskip 1em plus 0.5em minus 0.4em\relax PMLR, 2023, pp. 13\,916--13\,932.

\bibitem{kim2016future}
J.~Kim, ``The future of graphic and mobile memory for new applications,'' in \emph{2016 IEEE Hot Chips 28 Symposium (HCS)}.\hskip 1em plus 0.5em minus 0.4em\relax IEEE Computer Society, 2016, pp. 1--25.

\bibitem{kim2015ramulator}
Y.~Kim, W.~Yang, and O.~Mutlu, ``Ramulator: A fast and extensible dram simulator,'' \emph{IEEE Computer architecture letters}, vol.~15, no.~1, pp. 45--49, 2015.

\bibitem{kong2024cambricon}
W.~Kong, Y.~Hao, Q.~Guo, Y.~Zhao, X.~Song, X.~Li, M.~Zou, Z.~Du, R.~Zhang, C.~Liu \emph{et~al.}, ``Cambricon-d: Full-network differential acceleration for diffusion models,'' in \emph{2024 ACM/IEEE 51st Annual International Symposium on Computer Architecture (ISCA)}.\hskip 1em plus 0.5em minus 0.4em\relax IEEE, 2024, pp. 903--914.

\bibitem{lee2017understanding}
S.~Lee, Y.~Ro, Y.~H. Son, H.~Cho, N.~S. Kim, and J.~H. Ahn, ``Understanding power-performance relationship of energy-efficient modern dram devices,'' in \emph{2017 IEEE International Symposium on Workload Characterization (IISWC)}.\hskip 1em plus 0.5em minus 0.4em\relax IEEE, 2017, pp. 110--111.

\bibitem{li2023spada}
Z.~Li, J.~Li, T.~Chen, D.~Niu, H.~Zheng, Y.~Xie, and M.~Gao, ``Spada: Accelerating sparse matrix multiplication with adaptive dataflow,'' in \emph{Proceedings of the 28th ACM International Conference on Architectural Support for Programming Languages and Operating Systems, Volume 2}, 2023, pp. 747--761.

\bibitem{lu2022dpm}
C.~Lu, Y.~Zhou, F.~Bao, J.~Chen, C.~Li, and J.~Zhu, ``Dpm-solver++: Fast solver for guided sampling of diffusion probabilistic models,'' \emph{arXiv preprint arXiv:2211.01095}, 2022.

\bibitem{lu2021sanger}
L.~Lu, Y.~Jin, H.~Bi, Z.~Luo, P.~Li, T.~Wang, and Y.~Liang, ``Sanger: A co-design framework for enabling sparse attention using reconfigurable architecture,'' in \emph{MICRO-54: 54th Annual IEEE/ACM International Symposium on Microarchitecture}, 2021, pp. 977--991.

\bibitem{luo2023lcm}
S.~Luo, Y.~Tan, S.~Patil, D.~Gu, P.~von Platen, A.~Passos, L.~Huang, J.~Li, and H.~Zhao, ``Lcm-lora: A universal stable-diffusion acceleration module,'' \emph{arXiv preprint arXiv:2311.05556}, 2023.

\bibitem{mahmoud2018diffy}
M.~Mahmoud, K.~Siu, and A.~Moshovos, ``Diffy: A d{\'e}j{\`a} vu-free differential deep neural network accelerator,'' in \emph{2018 51st Annual IEEE/ACM International Symposium on Microarchitecture (MICRO)}.\hskip 1em plus 0.5em minus 0.4em\relax IEEE, 2018, pp. 134--147.

\bibitem{nvidia_diffusion_presentation_cvpr}
\BIBentryALTinterwordspacing
NVIDIA. Denoising diffusion-based generative modeling: Foundations and applications. [Online]. Available: \url{https://cvpr2022-tutorial-diffusion-models.github.io/}
\BIBentrySTDinterwordspacing

\bibitem{jetson_orin_nano}
\BIBentryALTinterwordspacing
NVIDIA. Jetson orin nano developer kit getting started guide. [Online]. Available: \url{https://developer.nvidia.com/embedded/learn/get-started-jetson-orin-nano-devkit}
\BIBentrySTDinterwordspacing

\bibitem{nvidia_a100}
\BIBentryALTinterwordspacing
NVIDIA. Nvidia a100 tensor core gpu. [Online]. Available: \url{https://www.nvidia.com/en-us/data-center/a100/}
\BIBentrySTDinterwordspacing

\bibitem{rtx6000}
\BIBentryALTinterwordspacing
NVIDIA. Nvidia rtx 6000 ada generation graphics card. [Online]. Available: \url{https://www.nvidia.com/en-us/design-visualization/rtx-6000/}
\BIBentrySTDinterwordspacing

\bibitem{nvidia_smi}
\BIBentryALTinterwordspacing
NVIDIA. System management interface smi. [Online]. Available: \url{https://developer.nvidia.com/system-management-interface}
\BIBentrySTDinterwordspacing

\bibitem{tegrastats_Utility}
\BIBentryALTinterwordspacing
NVIDIA. tegrastats utility. [Online]. Available: \url{https://docs.nvidia.com/drive/drive_os_5.1.6.1L/nvvib_docs/index.html#page/DRIVE_OS_Linux_SDK_Development_Guide/Utilities/util_tegrastats.html}
\BIBentrySTDinterwordspacing

\bibitem{pal2018outerspace}
S.~Pal, J.~Beaumont, D.-H. Park, A.~Amarnath, S.~Feng, C.~Chakrabarti, H.-S. Kim, D.~Blaauw, T.~Mudge, and R.~Dreslinski, ``Outerspace: An outer product based sparse matrix multiplication accelerator,'' in \emph{2018 IEEE International Symposium on High Performance Computer Architecture (HPCA)}.\hskip 1em plus 0.5em minus 0.4em\relax IEEE, 2018, pp. 724--736.

\bibitem{peebles2023scalable}
W.~Peebles and S.~Xie, ``Scalable diffusion models with transformers,'' in \emph{Proceedings of the IEEE/CVF International Conference on Computer Vision}, 2023, pp. 4195--4205.

\bibitem{qin2020sigma}
E.~Qin, A.~Samajdar, H.~Kwon, V.~Nadella, S.~Srinivasan, D.~Das, B.~Kaul, and T.~Krishna, ``Sigma: A sparse and irregular gemm accelerator with flexible interconnects for dnn training,'' in \emph{2020 IEEE International Symposium on High Performance Computer Architecture (HPCA)}.\hskip 1em plus 0.5em minus 0.4em\relax IEEE, 2020, pp. 58--70.

\bibitem{qin2023fact}
Y.~Qin, Y.~Wang, D.~Deng, Z.~Zhao, X.~Yang, L.~Liu, S.~Wei, Y.~Hu, and S.~Yin, ``Fact: Ffn-attention co-optimized transformer architecture with eager correlation prediction,'' in \emph{Proceedings of the 50th Annual International Symposium on Computer Architecture}, 2023, pp. 1--14.

\bibitem{qu2022dota}
Z.~Qu, L.~Liu, F.~Tu, Z.~Chen, Y.~Ding, and Y.~Xie, ``Dota: detect and omit weak attentions for scalable transformer acceleration,'' in \emph{Proceedings of the 27th ACM International Conference on Architectural Support for Programming Languages and Operating Systems}, 2022, pp. 14--26.

\bibitem{radford2021learning}
A.~Radford, J.~W. Kim, C.~Hallacy, A.~Ramesh, G.~Goh, S.~Agarwal, G.~Sastry, A.~Askell, P.~Mishkin, J.~Clark \emph{et~al.}, ``Learning transferable visual models from natural language supervision,'' in \emph{International conference on machine learning}.\hskip 1em plus 0.5em minus 0.4em\relax PMLR, 2021, pp. 8748--8763.

\bibitem{rombach2022high}
R.~Rombach, A.~Blattmann, D.~Lorenz, P.~Esser, and B.~Ommer, ``High-resolution image synthesis with latent diffusion models,'' in \emph{Proceedings of the IEEE/CVF conference on computer vision and pattern recognition}, 2022, pp. 10\,684--10\,695.

\bibitem{salimans2022progressive}
T.~Salimans and J.~Ho, ``Progressive distillation for fast sampling of diffusion models,'' \emph{arXiv preprint arXiv:2202.00512}, 2022.

\bibitem{sauer2022stylegan}
A.~Sauer, K.~Schwarz, and A.~Geiger, ``Stylegan-xl: Scaling stylegan to large diverse datasets,'' in \emph{ACM SIGGRAPH 2022 conference proceedings}, 2022, pp. 1--10.

\bibitem{shazeer2020gluvariantsimprovetransformer}
\BIBentryALTinterwordspacing
N.~Shazeer, ``Glu variants improve transformer,'' 2020. [Online]. Available: \url{https://arxiv.org/abs/2002.05202}
\BIBentrySTDinterwordspacing

\bibitem{song2023consistency}
Y.~Song, P.~Dhariwal, M.~Chen, and I.~Sutskever, ``Consistency models,'' \emph{arXiv preprint arXiv:2303.01469}, 2023.

\bibitem{srivastava2020matraptor}
N.~Srivastava, H.~Jin, J.~Liu, D.~Albonesi, and Z.~Zhang, ``Matraptor: A sparse-sparse matrix multiplication accelerator based on row-wise product,'' in \emph{2020 53rd Annual IEEE/ACM International Symposium on Microarchitecture (MICRO)}.\hskip 1em plus 0.5em minus 0.4em\relax IEEE, 2020, pp. 766--780.

\bibitem{synopsys_design_complier}
\BIBentryALTinterwordspacing
Synopsys. synopsys design complier. [Online]. Available: \url{https://www.synopsys.com/implementation-and-signoff/rtl-synthesis-test/dc-ultra.html}
\BIBentrySTDinterwordspacing

\bibitem{rtx6000area}
\BIBentryALTinterwordspacing
Techpowerup. Rtx 6000 ada generation specs. [Online]. Available: \url{https://www.techpowerup.com/gpu-specs/rtx-6000-ada-generation.c3933}
\BIBentrySTDinterwordspacing

\bibitem{tevet2023human}
\BIBentryALTinterwordspacing
G.~Tevet, S.~Raab, B.~Gordon, Y.~Shafir, D.~Cohen-or, and A.~H. Bermano, ``Human motion diffusion model,'' in \emph{The Eleventh International Conference on Learning Representations}, 2023. [Online]. Available: \url{https://openreview.net/forum?id=SJ1kSyO2jwu}
\BIBentrySTDinterwordspacing

\bibitem{tseng2023edge}
J.~Tseng, R.~Castellon, and K.~Liu, ``Edge: Editable dance generation from music,'' in \emph{Proceedings of the IEEE/CVF Conference on Computer Vision and Pattern Recognition}, 2023, pp. 448--458.

\bibitem{vaswani2017attention}
A.~Vaswani, N.~Shazeer, N.~Parmar, J.~Uszkoreit, L.~Jones, A.~N. Gomez, {\L}.~Kaiser, and I.~Polosukhin, ``Attention is all you need,'' \emph{Advances in neural information processing systems}, vol.~30, 2017.

\bibitem{wallace1964suggestion}
C.~S. Wallace, ``A suggestion for a fast multiplier,'' \emph{IEEE Transactions on electronic Computers}, no.~1, pp. 14--17, 1964.

\bibitem{wang2021spatten}
H.~Wang, Z.~Zhang, and S.~Han, ``Spatten: Efficient sparse attention architecture with cascade token and head pruning,'' in \emph{2021 IEEE International Symposium on High-Performance Computer Architecture (HPCA)}.\hskip 1em plus 0.5em minus 0.4em\relax IEEE, 2021, pp. 97--110.

\bibitem{wang2022energy}
Y.~Wang, Y.~Qin, D.~Deng, J.~Wei, Y.~Zhou, Y.~Fan, T.~Chen, H.~Sun, L.~Liu, S.~Wei \emph{et~al.}, ``An energy-efficient transformer processor exploiting dynamic weak relevances in global attention,'' \emph{IEEE Journal of Solid-State Circuits}, vol.~58, no.~1, pp. 227--242, 2022.

\bibitem{yangtrapezoid}
Y.~Yang, J.~S. Emer, and D.~Sanchez, ``Trapezoid: A versatile accelerator for dense and sparse matrix multiplications,'' in \emph{Proceedings of the 51th Annual International Symposium on Computer Architecture}.\hskip 1em plus 0.5em minus 0.4em\relax IEEE, 2024, pp. 931--945.

\bibitem{zhang2021gamma}
G.~Zhang, N.~Attaluri, J.~S. Emer, and D.~Sanchez, ``Gamma: Leveraging gustavson’s algorithm to accelerate sparse matrix multiplication,'' in \emph{Proceedings of the 26th ACM International Conference on Architectural Support for Programming Languages and Operating Systems}, 2021, pp. 687--701.

\bibitem{zhang2020sparch}
Z.~Zhang, H.~Wang, S.~Han, and W.~J. Dally, ``Sparch: Efficient architecture for sparse matrix multiplication,'' in \emph{2020 IEEE International Symposium on High Performance Computer Architecture (HPCA)}.\hskip 1em plus 0.5em minus 0.4em\relax IEEE, 2020, pp. 261--274.

\bibitem{zheng2023fasttrain}
H.~Zheng, W.~Nie, A.~Vahdat, and A.~Anandkumar, ``Fast training of diffusion models with masked transformers,'' \emph{arXiv preprint arXiv:2306.09305}, 2023.

\bibitem{zheng2023fast}
H.~Zheng, W.~Nie, A.~Vahdat, K.~Azizzadenesheli, and A.~Anandkumar, ``Fast sampling of diffusion models via operator learning,'' in \emph{International conference on machine learning}.\hskip 1em plus 0.5em minus 0.4em\relax PMLR, 2023, pp. 42\,390--42\,402.

\end{thebibliography}
\end{document}